\documentstyle{arPrep}
\begin{document}

\input epsf.def   
\input psfig.sty  

\jname{Annu. Rev. Astron. Astrophys.}
\jyear{2000}
\jvol{}

%***
%***Definitions
%***

\newcommand\lya{Ly$\alpha$~}
\newcommand\ljeans{\lambda_{\rm J}}
\newcommand\ie{i.e.~}
\newcommand\kjeans{k_{\rm J}}
\newcommand\mjeans{M_{\rm J}}
\newcommand\beq{\begin{equation}}
\newcommand\eeq{\end{equation}}
\newcommand\Msun{$M_\odot$}
\newcommand\lyasource{{\dot N_\alpha}}
\newcommand\beqa{\begin{eqnarray}}
\newcommand\eeqa{\end{eqnarray}}
\newcommand\xb{{\bf x}}
\newcommand\rb{{\bf r}}
\newcommand\vb{{\bf v}}
\newcommand\ub{{\bf u}}
\newcommand\kb{{\bf k}}
\newcommand\Omm{{\Omega_m}}
\newcommand\Ommz{{\Omega_m^{\,z}}}
\newcommand\Omr{{\Omega_r}}
\newcommand\Omk{{\Omega_k}}
\newcommand\Oml{{\Omega_{\Lambda}}}
\newcommand\nb{\bar{n}}
\newcommand\etal{et al.\ }
\newcommand\Ng{N_\gamma}
\newcommand\zr{z_{\rm reion}}
\newcommand\HI{\rm H\,I}
\newcommand\cN{c_{\rm N}}
\newcommand\kB{k_{\rm B}}
\newcommand\Ni{N_{\rm ion}}
\newcommand\fg{f_{\rm gas}}
\newcommand\fe{f_{\rm eject}}
\newcommand\fin{f_{\rm int}}
\newcommand\fw{f_{\rm wind}}
\newcommand\NGST{{\it NGST}}
\newcommand\la{\mathrel{\hbox{\rlap{\hbox{\lower4pt\hbox{$\sim$}}}\hbox{$<$}}}}
\newcommand\ga{\mathrel{\hbox{\rlap{\hbox{\lower4pt\hbox{$\sim$}}}\hbox{$>$}}}}
\def\ionAR#1#2{#1$\;${\small\rm #2}\relax}
\newcommand\sun{\hbox{$\odot$}}
\newcommand\farcs{\hbox{$.\!\!^{\prime\prime}$}}
\newcommand\arcmin{\hbox{$^\prime$}}

%***
%***Title/Author
%***

\title{\bf The Reionization of the Universe by the First Stars and
Quasars}

\author{Abraham Loeb \affiliation{Department of Astronomy, Harvard
University, 60 Garden St., Cambridge, MA 02138; aloeb@cfa.harvard.edu}
Rennan Barkana \affiliation{Institute for Advanced Study, Olden Lane,
Princeton, NJ 08540; \\{\it Present address:} CITA, 60 St. George Street,
Toronto, Ontario, M5S 3H8, CANADA; barkana@cita.utoronto.ca } }

\markboth{Loeb \& Barkana}{Reionization of the Universe}

%***
%***Abstract
%***

\begin{abstract}

The formation of the first stars and quasars marks the transformation
of the universe from its smooth initial state to its clumpy current
state. In popular cosmological models, the first sources of light
began to form at a redshift $z=30$ and reionized most of the hydrogen
in the universe by $z=7$. Current observations are at the threshold of
probing the hydrogen reionization epoch. The study of high-redshift
sources is likely to attract major attention in observational and
theoretical cosmology over the next decade.
\vskip 0.5truein
{\it Key
Words: Cosmology, First Galaxies, Intergalactic Medium}
%\vfill\eject
\vspace{.4in}
\end{abstract}

\maketitle

%\vfill\eject
\vspace{.5in}

%<>**************************************************************************

\section{PREFACE: THE FRONTIER OF SMALL-SCALE STRUCTURE}
\label{sec1}

The detection of cosmic microwave background (CMB) anisotropies
(Bennett et al.\  1996; de Bernardis et al.\  2000; Hanany et al.\  2000)
confirmed the notion that the present large-scale structure in the
universe originated from small-amplitude density fluctuations at early
times.
%The amplitude of the inferred density fluctuations 
%is compatible with the values necessary to account  
%for the large-scale structures of the 
%distribution of galaxies in the local universe (cite).
%Extrapolation of the fluctuation spectrum to small scales
%in the context of inflationary scale-invariant initial 
%conditions and cold dark matter gives the correct
%abundance of X-ray clusters and galaxies.
Due to the natural instability of gravity, regions that were denser than
average collapsed and formed bound objects, first on small spatial
scales and later on larger and larger scales. The present-day
abundance of bound objects, such as galaxies and X-ray clusters, can be
explained based on an appropriate extrapolation of the detected
anisotropies to smaller scales.
%(e.g., Baugh et al.\  1997; cite example for clusters). 
Existing observations with the {\it Hubble Space Telescope}\, (e.g.,
Steidel et al.\ 1996; Madau et al.\ 1996; Chen et al.\ 1999; Clements
et al.\ 1999) and ground-based telescopes (Lowenthal et al.\ 1997; Dey
et al.\ 1999; Hu et al.\ 1998, 1999; Spinrad et al.\ 1999; Steidel et
al.\ 1999), have constrained the evolution of galaxies and their
stellar content at $z\la 6$.  However, in the bottom-up hierarchy of
the popular Cold Dark Matter (CDM) cosmologies, galaxies were
assembled out of building blocks of smaller mass. The elementary
building blocks, i.e., the first gaseous objects to form, acquired a
total mass of order the Jeans mass ($\sim 10^4 M_\odot$), below which
gas pressure opposed gravity and prevented collapse (Couchman \& Rees
1986; Haiman \& Loeb 1997; Ostriker \& Gnedin 1997). In variants of
the standard CDM model, these basic building blocks first formed at
$z\sim 15$--$30$.

\begin{figure}
\psfig{file=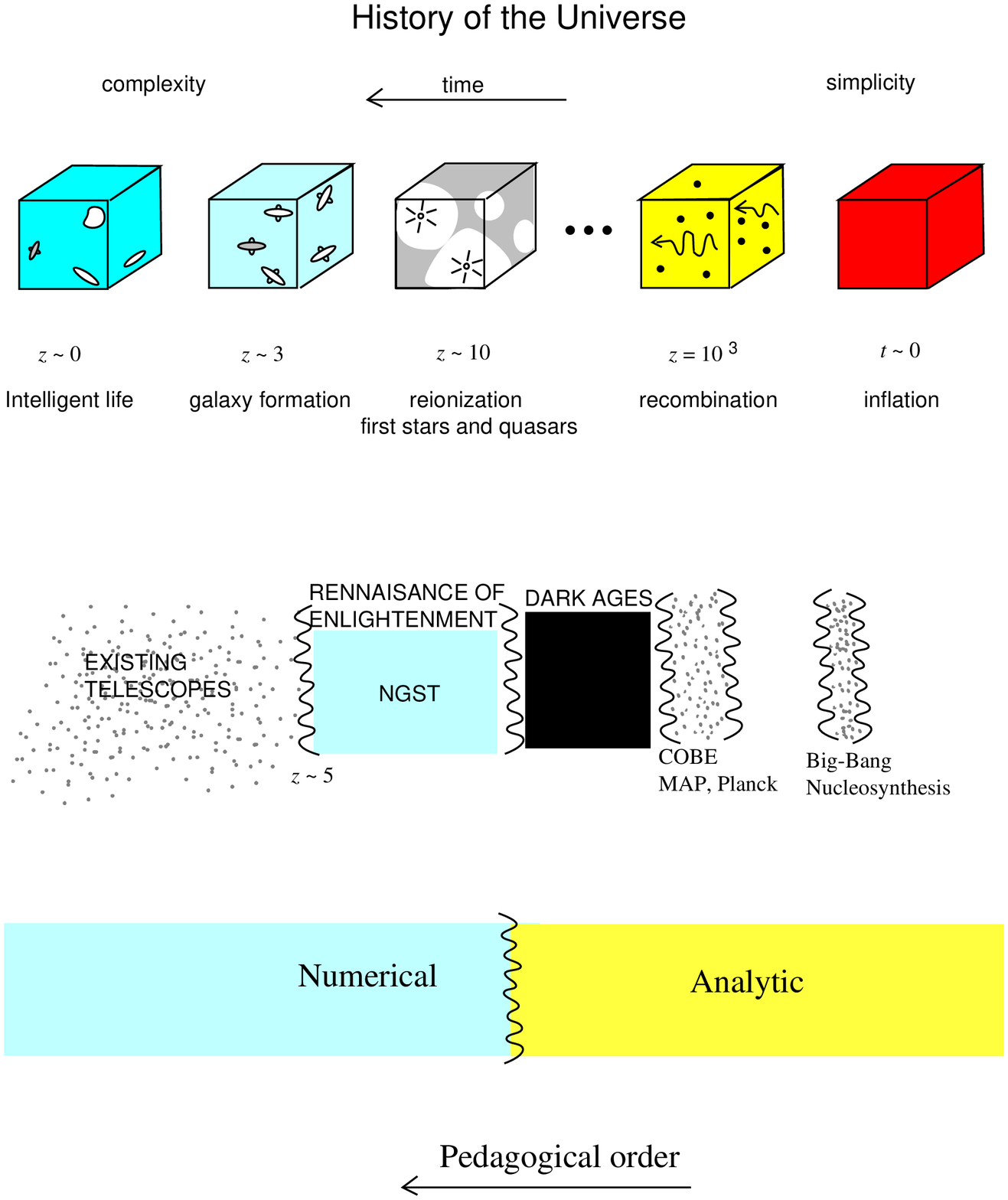,width=4.8in}
\caption{Milestones in the evolution of the universe from simplicity
to complexity. The ``end of the dark ages'' bridges between the
recombination epoch probed by microwave anisotropy experiments ($z\sim
10^3$) and the horizon of current observations ($z\sim 5$--6).}
\label{fig1a}
\end{figure}

An important qualitative outcome of the microwave anisotropy data is
the confirmation that the universe started out simple. It was by and
large homogeneous and isotropic with small fluctuations that can be
described by linear perturbation analysis. The current universe is
clumpy and complicated.  Hence, the arrow of time in cosmic history
also describes the progression from simplicity to complexity (see
Figure~\ref{fig1a}). While the conditions in the early universe can be
summarized on a single sheet of paper, the mere description of the
physical and biological structures found in the present-day universe
cannot be captured by thousands of books in our libraries.  The
formation of the first bound objects marks the central milestone in
the transition from simplicity to complexity.  Pedagogically, it would
seem only natural to attempt to understand this epoch before we try to
explain the present-day universe. Historically, however, most of the
astronomical literature focused on the local universe and has only
been shifting recently to the early universe. This violation of
the pedagogical rule was forced upon us by the limited state of our
technology; observation of earlier cosmic times requires detection of
distant sources, which is feasible only with large telescopes and
highly-sensitive instrumentation.

For these reasons, advances in technology are likely to make the high
redshift universe an important frontier of cosmology over the coming
decade. This effort will involve large (30 meter) ground-based
telescopes and will culminate in the launch of the successor to the
{\it Hubble Space Telescope},\, called {\it Next Generation Space
Telescope}\footnote{More details about the telescope can be found at
http://ngst.gsfc.nasa.gov/} (\NGST\,). \NGST,\, planned for launch in
2009, will image the first sources of light that formed in the
universe. With its exceptional sub-nJy ($1~{\rm nJy}=10^{-32}{\rm
erg~cm^{-2}~s^{-1}~Hz^{-1}}$) sensitivity in the 1--3.5$\mu$m infrared
regime, \NGST\, is ideally suited for probing optical-UV emission from
sources at redshifts $\ga 10$, just when popular Cold Dark Matter
models for structure formation predict the first baryonic objects to
have collapsed.

The study of the the formation of the first generation of sources at early
cosmic times (high redshifts) holds the key to constraining the
power-spectrum of density fluctuations on small scales. Previous
research in cosmology has been dominated by studies of {\it Large Scale
Structure} (LSS); future studies are likely to focus on {\it Small
Scale Structure} (SSS).

The first sources are a direct consequence of the growth of linear
density fluctuations. As such, they emerge from a well-defined set of
initial conditions and the physics of their formation can be followed
precisely by computer simulation. The cosmic initial conditions for
the formation of the first generation of stars are much simpler than
those responsible for star formation in the Galactic interstellar
medium at present. The cosmic conditions are fully specified by the
primordial power spectrum of Gaussian density fluctuations, the mean
density of dark matter, the initial temperature and density of the
cosmic gas, the primordial composition according to Big-Bang
nucleosynthesis, and the lack of dynamically-significant magnetic
fields. The chemistry is much simpler in the absence of metals and the
gas dynamics is much simpler in the absence of dynamically-important
magnetic fields.

The initial mass function of the first stars and black holes is therefore
determined by a simple set of initial conditions (although subsequent
generations of stars are affected by feedback from photoionization heating
and metal enrichment).  While the early evolution of the seed density
fluctuations can be fully described analytically, the collapse and
fragmentation of nonlinear structure must be simulated numerically. The
first baryonic objects connect the simple initial state of the universe to
its complex current state, and their study with hydrodynamic simulations
(e.g., Abel, Bryan, \& Norman 2000; Bromm, Coppi, \& Larson 1999) and with
future telescopes such as \NGST\, offers the key to advancing our knowledge
on the formation physics of stars and massive black holes.

\begin{figure}
\psfig{file=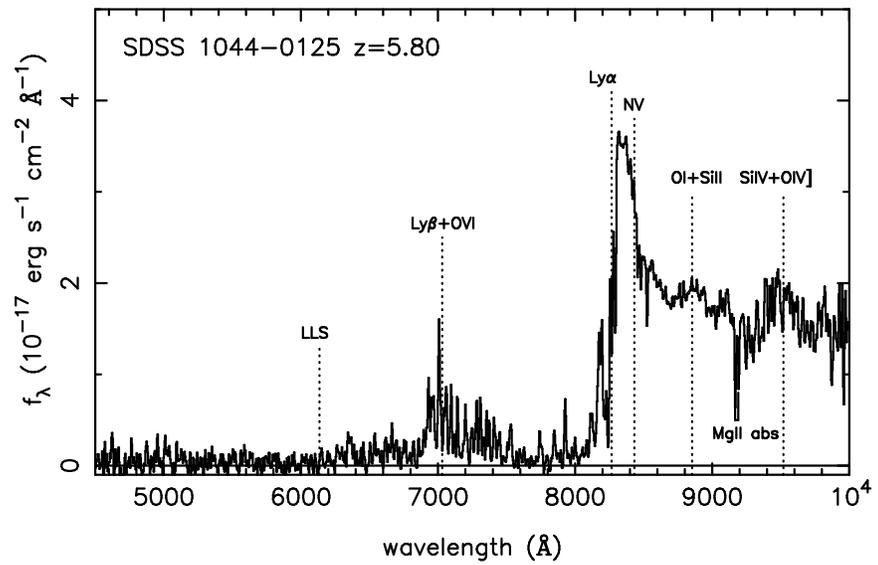,width=5in}
\caption{Optical spectrum of the highest-redshift known quasar at
$z=5.8$, discovered by the Sloan Digital Sky Survey (Fan et
al.\  2000). }
\label{fig1c}
\end{figure}

The {\it first light} from stars and quasars ended the ``dark ages''
\footnote{The use of this term in the cosmological context was coined
by Sir Martin Rees.} of the universe and initiated a ``renaissance of
enlightenment'' in the otherwise fading glow of the microwave
background (see Figure~\ref{fig1a}).  It is easy to see why the mere
conversion of trace amounts of gas into stars or black holes at this
early epoch could have had a dramatic effect on the ionization state
and temperature of the rest of the gas in the universe. Nuclear fusion
releases $\sim 7\times 10^6$ eV per hydrogen atom, and thin-disk
accretion onto a Schwarzschild black hole releases ten times more
energy; however, the ionization of hydrogen requires only 13.6 eV. It
is therefore sufficient to convert a small fraction, $\sim 10^{-5}$ of
the total baryonic mass into stars or black holes in order to ionize
the rest of the universe. (The actual required fraction is higher by
at least an order of magnitude [Bromm, Kudritzky, \& Loeb 2000]
because only some of the emitted photons are above the ionization
threshold of 13.6 eV and because each hydrogen atom recombines more
than once at redshifts $z\ga 7$). Recent calculations of structure
formation in popular CDM cosmologies imply that the universe was
ionized at $z\sim 7$--12 (Haiman \& Loeb 1998, 1999b,c; Gnedin \&
Ostriker 1998; Chiu \& Ostriker 2000; Gnedin 2000a). Current
observations are at the threshold of probing this epoch of
reionization, given the fact that galaxies and quasars at redshifts
$\sim 6$ are being discovered (Fan et al.\ 2000; Stern et al.\
2000). One of these sources is a bright quasar at $z=5.8$ whose
spectrum is shown in Figure~\ref{fig1c}. The plot indicates that there
is transmitted flux shortward of the Ly$\alpha$ wavelength at the
quasar redshift. However, Gunn \& Peterson (1965) showed that even a
tiny neutral hydrogen fraction in the intergalactic medium would
produce a large optical depth at these wavelengths. Indeed, modeling
of the transmitted flux (Fan et al.\ 2000) implies an optical depth
$\tau<0.5$ or a neutral fraction $x_{\HI}\la 10^{-6}$, i.e., the
universe is fully ionized at $z=5.8$!  One of the important challenges
of future observations will be to identify {\it when and how the
intergalactic medium was ionized}. Theoretical calculations (see \S
\ref{sec6.3.1}) imply that such observations are just around the
corner.

\begin{figure}
\psfig{file=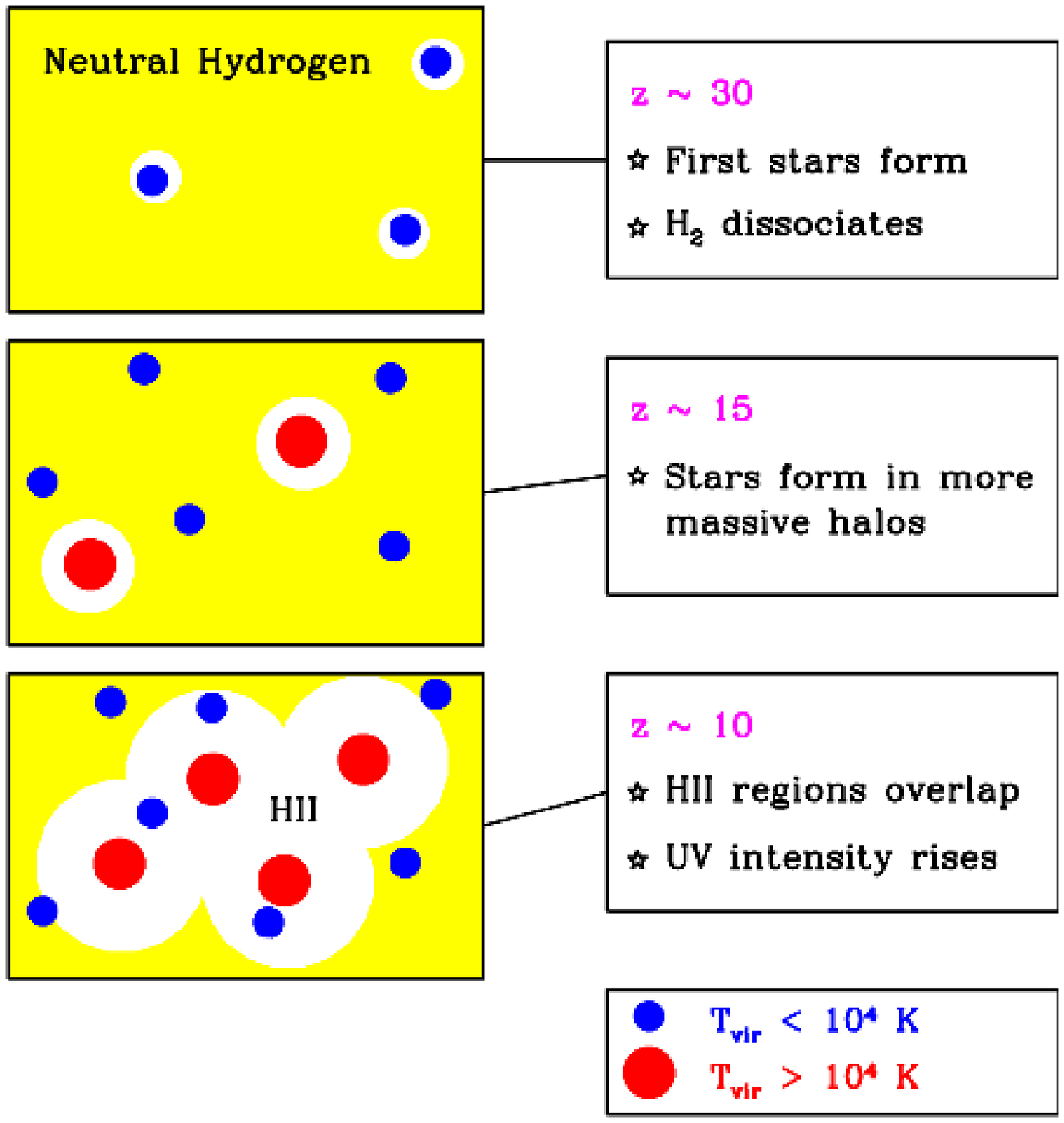,width=6in}
\caption{Stages in the reionization of hydrogen in the intergalactic
medium. }
\label{fig1d}
\end{figure}

Figure~\ref{fig1d} shows schematically the various stages in a
theoretical scenario for the history of hydrogen reionization in the
intergalactic medium. The first gaseous clouds collapse at redshifts
$\sim 20$--$30$ and fragment into stars due to molecular hydrogen
(H$_2$) cooling. However, H$_2$ is fragile and can be easily
dissociated by a small flux of UV radiation (Stecher \& Williams 1967;
Haiman, Rees, \& Loeb 1997).  Hence the bulk of the radiation that
ionized the universe is emitted from galaxies with a virial
temperature $\ga 10^4$ K, where atomic cooling is effective and allows
the gas to fragment (see Haiman, Abel, \& Rees 1999 for an alternative
scenario). 

Since recent observations confine the standard set of cosmological
parameters to a relatively narrow range, we assume a $\Lambda$CDM
cosmology with a particular standard set of parameters in the
quantitative results in this review. For the contributions to the
energy density, we assume ratios relative to the critical density of
$\Omm=0.3$, $\Oml=0.7$, and $\Omega_b=0.045$, for matter, vacuum
(cosmological constant), and baryons, respectively. We also assume a
Hubble constant $H_0=70\mbox{ km s}^{-1}\mbox{Mpc}^{-1}$, and a
primordial scale invariant ($n=1$) power spectrum with $\sigma_8=0.9$,
where $\sigma_8$ is the root-mean-square amplitude of mass
fluctuations in spheres of radius $8\ h^{-1}$ Mpc. These parameter
values are based primarily on the following observational results: CMB
temperature anisotropy measurements on large scales (Bennett et al.\
1996) and on the scale of $\sim 1^\circ$ (Lange et al.\ 2000; Balbi et
al.\ 2000); the abundance of galaxy clusters locally (Viana \& Liddle
1999; Pen 1998; Eke, Cole, \& Frenk 1996) and as a function of
redshift (Bahcall \& Fan 1998; Eke, Cole, Frenk, \& Henry 1998); the
baryon density inferred from big bang nucleosynthesis (see the review
by Tytler et al.\ 2000); distance measurements used to derive the
Hubble constant (Mould et al.\ 2000; Jha et al.\ 1999; Tonry et al.\
1997); and indications of cosmic acceleration from distances based on
type Ia supernovae (Perlmutter et al.\ 1999; Riess et al.\ 1998).

This review summarizes recent theoretical advances in understanding
the physics of the first generation of cosmic structures. Although the
literature on this subject extends all the way back to the sixtees
(Saslaw \& Zipoy 1967, Peebles \& Dicke 1968, Hirasawa 1969, Matsuda
et al.\ 1969, Hutchins 1976, Silk 1983, Palla et al.\ 1983, Lepp \&
Shull 1984, Couchman 1985, Couchman \& Rees 1986, Lahav 1986), this
review focuses on the progress made over the past decade in the modern
context of CDM cosmologies.

%Consider adding images of HDF ($z\la 6$) and COBE ($z=10^3$)

%<>***************************************************************************

\section{RADIATIVE FEEDBACK FROM THE FIRST SOURCES OF LIGHT}
\label{sec6}

\subsection{Escape of Ionizing Radiation from Galaxies}
\label{sec6.1}

The intergalactic ionizing radiation field, a key ingredient in the
development of reionization, is determined by the amount of ionizing
radiation escaping from the host galaxies of stars and quasars.  The
value of the escape fraction as a function of redshift and galaxy mass
remains a major uncertainty in all current studies, and could affect
the cumulative radiation intensity by orders of magnitude at any given
redshift. Gas within halos is far denser than the typical density of
the IGM, and in general each halo is itself embedded within an
overdense region, so the transfer of the ionizing radiation must be
followed in the densest regions in the universe. Reionization
simulations are limited in resolution and often treat the sources of
ionizing radiation and their immediate surroundings as unresolved
point sources within the large-scale intergalactic medium (see, e.g.,
Gnedin 2000a).

The escape of ionizing radiation ($h\nu > 13.6$eV, $\lambda < 912$
{\AA}) from the disks of present-day galaxies has been studied in
recent years in the context of explaining the extensive diffuse
ionized gas layers observed above the disk in the Milky Way (Reynolds
et al.\ 1995) and other galaxies (e.g., Rand 1996; Hoopes, Walterbos,
\& Rand 1999). Theoretical models predict that of order 3--14\% of the
ionizing luminosity from O and B stars escapes the Milky Way disk
(Dove \& Shull 1994; Dove, Shull, \& Ferrara 2000).  A similar escape
fraction of $f_{\rm esc}=6$\% was determined by Bland-Hawthorn \&
Maloney (1999) based on H$\alpha$ measurements of the Magellanic
Stream.  From {\it Hopkins Ultraviolet Telescope} observations of four
nearby starburst galaxies (Leitherer et al.\ 1995; Hurwitz, Jelinsky,
\& Dixon 1997), the escape fraction was estimated to be in the range
3\%$<f_{\rm esc} < 57$\%.  If similar escape fractions characterize
high redshift galaxies, then stars could have provided a major
fraction of the background radiation that reionized the IGM (e.g.,
Madau \& Shull 1996; Madau 1999).  However, the escape fraction from
high-redshift galaxies, which formed when the universe was much denser
($\rho\propto (1+z)^3$), may be significantly lower than that
predicted by models ment to describe present-day galaxies.  Current
reionization calculations assume that galaxies are isotropic point
sources of ionizing radiation and adopt escape fractions in the range
$5\% < f_{\rm esc} < 60\%$ (see, e.g., Gnedin 2000a, Miralda-Escud\'e
et al.\ 1999).

Clumping is known to have a significant effect on the penetration and
escape of radiation from an inhomogeneous medium (e.g., Boiss\'e 1990;
Witt \& Gordon 1996, 2000; Neufeld 1991; Haiman \& Spaans 1999;
Bianchi et al.\ 2000).  The inclusion of clumpiness introduces several
unknown parameters into the calculation, such as the number and
overdensity of the clumps, and the spatial correlation between the
clumps and the ionizing sources.  An additional complication may arise
from hydrodynamic feedback, whereby part of the gas mass is expelled
from the disk by stellar winds and supernovae [see \S 7 of Barkana \&
Loeb (2000c) for a review of this topic].

\begin{figure} 
\psfig{file=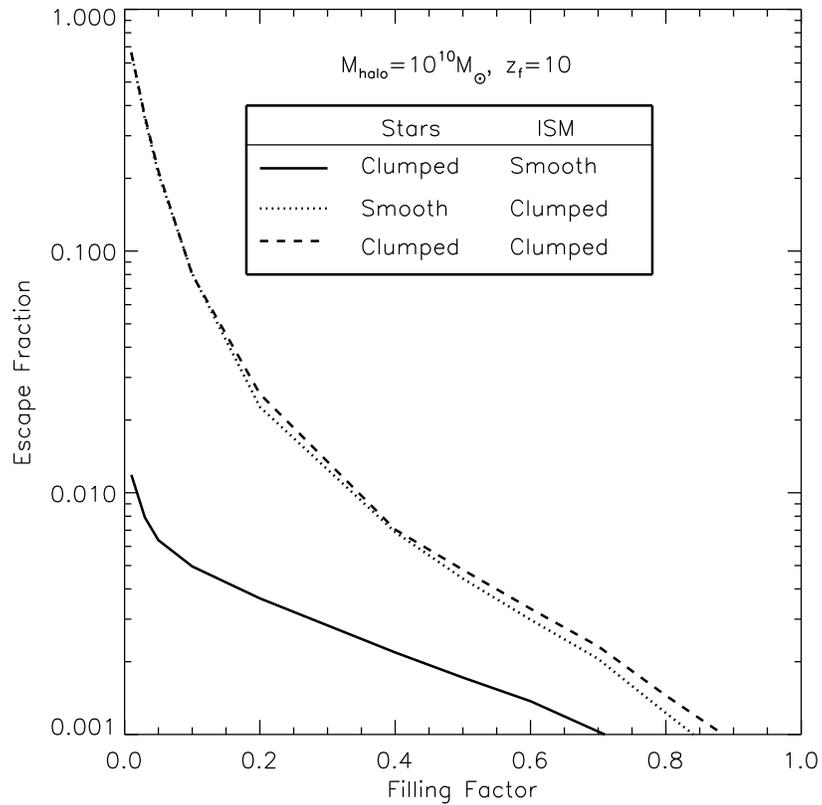,width=9in}
\caption{Escape fractions of stellar ionizing photons from a gaseous
disk embedded within a $10^{10}M_\odot$ halo which have formed at
$z=10$ (from Wood \& Loeb 2000). The curves show three different cases
of clumpiness within the disk. The volume filling factor refers to
either the ionizing emissivity, the gas clumps, or both, depending on
the case. The escape fraction is substantial ($\ga 1\%$) only if the
gas distribution is highly clumped.  }
\label{fig6a}
\end{figure}

Wood \& Loeb (2000) used a three-dimensional radiation transfer code
to calculate the steady-state escape fraction of ionizing photons from
disk galaxies as a function of redshift and galaxy mass. The gaseous
disks were assumed to be isothermal, with a sound speed $c_s\sim
10~{\rm km~s^{-1}}$, and radially exponential. For stellar sources,
the predicted increase in the disk density with redshift resulted in a
strong decline of the escape fraction with increasing redshift. The
situation is different for a central quasar. Due to its higher
luminosity and central location, the quasar tends to produce an
ionization channel in the surrounding disk through which much of its
ionizing radiation escapes from the host. In a steady state, only
recombinations in this ionization channel must be balanced by
ionizations, while for stars there are many ionization channels
produced by individual star-forming regions and the total
recombination rate in these channels is very high. Escape fractions
$\ga 10\%$ were achieved for stars at $z\sim 10$ only if $\sim 90\%$
of the gas was expelled from the disks or if dense clumps removed the
gas from the vast majority ($\ga 80\%$) of the disk volume (see
Figure~\ref{fig6a}). This analysis applies only to halos with virial
temperatures $\ga 10^4$ K. Ricotti \& Shull (2000) reached similar
conclusions but for a quasi-spherical configuration of stars and gas.
They demonstrated that the escape fraction is substantially higher in
low-mass halos with a virial temperature $\la 10^4$ K.  However, the
formation of stars in such halos depends on their uncertain ability to
cool via the efficient production of molecular hydrogen (Haiman, Rees,
\& Loeb 1997; Haiman, Abel, \& Rees 1999).

\noindent
  
The main uncertainty in the above predictions involves the
distribution of the gas inside the host galaxy, as the gas is exposed
to the radiation released by stars and the mechanical energy deposited
by supernovae.  Given the fundamental role played by the escape
fraction, it is desirable to calibrate its value observationally.
Recently, Steidel, Pettini, \& Adelberger (2000) reported a
preliminary detection of significant Lyman continuum flux in the
composite spectrum of 29 Lyman break galaxies (LBG) with redshifts in
the range $z = 3.40\pm 0.09$. After correcting for intergalactic
absorption, Steidel et al.\ (2000) inferred a ratio between the
emergent flux density at 1500\AA~ and 900\AA~ (rest frame) of $4.6 \pm
1.0$. Taking into account the fact that the stellar spectrum should
already have an intrinsic Lyman discontinuity of a factor of $\sim
3$--5, but that only $\sim 15$--$20\%$ of the 1500\AA~ photons escape
from typical LBGs without being absorbed by dust (Pettini et al.\
1998a; Adelberger \& Steidel 2000), the inferred 900\AA~ escape
fraction is $f_{\rm esc} \sim 10$--$20\%$.  Although the galaxies in
this sample were drawn from the bluest quartile of the LBG spectral
energy distributions, the measurement implies that this quartile may
itself dominate the hydrogen-ionizing background relative to quasars
at $z\sim 3$.

%<>
\subsection{Propagation of Ionization Fronts in the IGM}
\label{sec6.2}

The radiation output from the first stars ionizes hydrogen in a
growing volume, eventually encompassing almost the entire IGM within a
single \ionAR{H}{II} bubble. In the early stages of this process, each
galaxy produces a distinct \ionAR{H}{II} region, and only when the
overall \ionAR{H}{II} filling factor becomes significant do neighboring
bubbles begin to overlap in large numbers, ushering in the ``overlap
phase'' of reionization. Thus, the first goal of a model of
reionization is to describe the initial stage, when each source
produces an isolated expanding \ionAR{H}{II} region.

We assume a spherical ionized volume which is separated from the
surrounding neutral gas by a sharp ionization front. Indeed, in the
case of a stellar ionizing spectrum, most ionizing photons are just
above the hydrogen ionization threshold of 13.6 eV, where the
absorption cross-section is high and a very thin layer of neutral
hydrogen is sufficient to absorb all the ionizing photons. On the
other hand, an ionizing source such as a quasar produces significant
numbers of higher energy photons and results in a thicker transition
region.

In the absence of recombinations, each hydrogen atom in the IGM would
only have to be ionized once. However, the increased density of the
IGM at high redshift implies that recombinations cannot be neglected,
since the recombination rate depends on the square of the gas
density. Thus, if the IGM is not uniform, but instead the gas which is
being ionized is mostly distributed in high-density clumps, then the
recombination rate is very high. This is often dealt with by
introducing a volume-averaged clumping factor $C$ (in general
time-dependent), defined by\footnote{The recombination rate depends on
the number density of electrons, and in using equation~(\ref{clump})
we are neglecting the small contribution caused by partially or fully
ionized helium.} \beq C=\left<n_H^2\right>/\nb_H^2 \label{clump}\
. \eeq If the ionized volume is large compared to the typical scale of
clumping, so that many clumps are averaged over, then the solution for
the comoving volume $V(t)$ (generalized from Shapiro \& Giroux 1987;
see Barkana \& Loeb 2000c) around a source which turns on at $t=t_i$
is \beq V(t)=\int_{t_i}^t \frac{1}{\nb_H^0} \frac{d\, \Ng}{dt'}
\,e^{F(t',t)} dt'\ ,\label{HIIsoln} \eeq where \beq F(t',t)=-\alpha_B
\nb_H^0 \int_{t'}^t \frac{C(t'')} {a^3(t'')}\, dt''\
\label{Fgen}. \eeq In these expressions, $d \Ng/dt$ is the rate at
which the source produces ionizing photons, $\nb_H^0$ is the present
number density of hydrogen, $a$ is the scale factor, and $\alpha_B$ is
the case B recombination coefficient for hydrogen at $T=10^4$ K. 

The size of the resulting \ionAR{H}{II} region depends on the halo
which produces it. Consider a halo of total mass $M$ and baryon
fraction $\Omega_b/\Omm$. To derive a rough estimate, we assume that
baryons are incorporated into stars with an efficiency of $f_{\rm
star}=10\%$, and that the escape fraction for the resulting ionizing
radiation is also $f_{\rm esc}=10\%$. If the stellar IMF is similar to
the one measured locally (Scalo 1998), then $N_{\gamma} \approx 4000$
ionizing photons are produced per baryon in stars (for a metallicity
equal to $1/20$ of the solar value). We define a parameter which gives
the overall number of ionizations per baryon, \beq \Ni \equiv
N_{\gamma} \, f_{\rm star}\, f_{\rm esc}\ . \eeq If we neglect
recombinations then we obtain the maximum comoving radius of the
region which the halo of mass $M$ can ionize, \beq r_{\rm max}=
\left(\frac{3}{4\pi}\, \frac{\Ng} {\nb_H^0} \right)^{1/3} =
\left(\frac{3}{4\pi}\, \frac{\Ni} {\nb_H^0}\, \frac{\Omega_b}{\Omm}\,
\frac{M}{m_p} \right)^{1/3}= 680\, {\rm kpc} \left( \frac{\Ni}{40}\,
\frac{M} {10^8 M_{\sun}}\right)^{1/3}\ ,
\label{rmax} \eeq for our standard choice of cosmological parameters:
$\Omm=0.3$, $\Oml=0.7$, and $\Omega_b=0.045$. The actual radius never
reaches this size if the recombination time is shorter than the
lifetime of the ionizing source.

\begin{figure}
\psfig{file=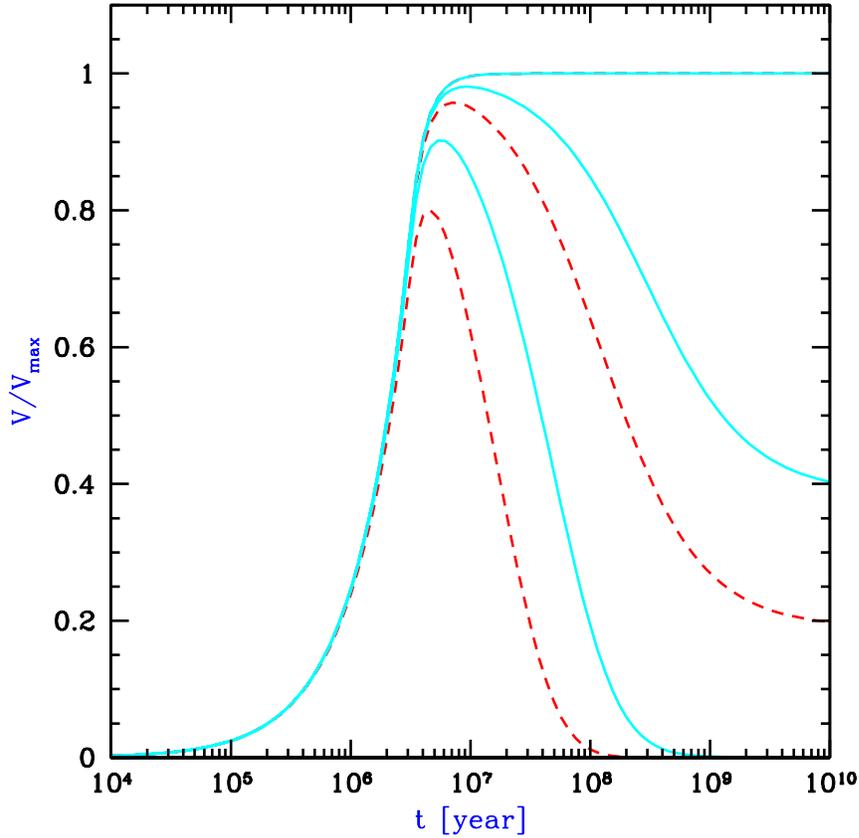,width=4.7in}
%Used FHIIbub.mon
\caption{Expanding \ionAR{H}{II} region around an isolated ionizing
source. The comoving ionized volume $V$ is expressed in units of the
maximum possible volume, $V_{\rm max}=4\pi r_{\rm max}^3/3$ [with
$r_{\rm max}$ given in equation~(\ref{rmax})], and the time is
measured after an instantaneous starburst which is characterized by a
Scalo (1998) IMF. We consider a source turning on at $z=10$ (solid
curves) or $z=15$ (dashed curves), with three cases for each: no
recombinations, $C=1$, and $C=10$, in order from top to bottom. The
no-recombination curve is identical for the different source
redshifts.}
\label{fig6b}
\end{figure}

In Figure~\ref{fig6b} we show the time evolution of the volume ionized
by such a source which undergoes an instantaneous starburst with the
Scalo (1998) IMF. The volume is shown in units of the maximum volume
$V_{\rm max}$ which corresponds to $r_{\rm max}$ in
equation~(\ref{rmax}). We consider a source turning on at $z=10$
(solid curves) or $z=15$ (dashed curves), with three cases for each:
no recombinations, $C=1$, and $C=10$, in order from top to bottom
(Note that the result is independent of redshift in the case of no
recombinations; also, in every case we assume a time-independent
$C$). When recombinations are included, the volume rises and reaches
close to $V_{\rm max}$ before dropping after the source turns off. At
large $t$ recombinations stop due to the dropping density, and the
volume approaches a constant value (although $V \ll V_{\rm max}$ at
large $t$ if $C=10$).

%%%%%%%% Figure6.2
%%%%%%%%

We obtain a similar result for the size of the \ionAR{H}{II} region
around a galaxy if we consider a mini-quasar rather than stars. For
the typical quasar spectrum (Elvis et al.\ 1994), if we assume a
radiative efficiency of $\sim 6\%$ then roughly 11,000 ionizing
photons are produced per baryon incorporated into the black hole
(Haiman, personal communication). The efficiency of incorporating the
baryons in a galaxy into a central black hole is low ($\la 0.6\%$ in
the local universe, e.g.\ Magorrian et al.\ 1998; see also \S
\ref{sec8.2.2}), but the escape fraction for quasars is likely to be
close to unity, i.e., an order of magnitude higher than for stars (see
\S \ref{sec6.1}). Thus, for every baryon in galaxies, up to $\sim 65$
ionizing photons may be produced by a central black hole and $\sim 40$
by stars, although both of these numbers for $\Ni$ are highly
uncertain. These numbers suggest that in either case the typical size
of \ionAR{H}{II} regions before reionization may be $\la 1$ Mpc or
$\sim 10$ Mpc, depending on whether $10^8 M_{\sun}$ halos or $10^{12}
M_{\sun}$ halos dominate.

\subsection{Reionization of Hydrogen in the IGM}
\label{sec6.3.1}

In this section we summarize recent progress, both analytic and
numerical, made toward elucidating the basic physics of reionization
and the way in which the characteristics of reionization depend on the
nature of the ionizing sources and on other input parameters of
cosmological models.

The process of the reionization of hydrogen involves several distinct
stages. The initial, ``pre-overlap'' stage (using the terminology of
Gnedin 2000a) consists of individual ionizing sources turning on and
ionizing their surroundings. The first galaxies form in the most
massive halos at high redshift, and these halos are biased and are
preferentially located in the highest-density regions. Thus the
ionizing photons which escape from the galaxy itself (see \S
\ref{sec6.1}) must then make their way through the surrounding
high-density regions, which are characterized by a high recombination
rate. Once they emerge, the ionization fronts propagate more easily
into the low-density voids, leaving behind pockets of neutral,
high-density gas. During this period the IGM is a two-phase medium
characterized by highly ionized regions separated from neutral regions
by ionization fronts.  Furthermore, the ionizing intensity is very
inhomogeneous even within the ionized regions, with the intensity
determined by the distance from the nearest source and by the ionizing
luminosity of this source.

The central, relatively rapid ``overlap'' phase of reionization begins when
neighboring \ionAR{H}{II} regions begin to overlap. Whenever two ionized
bubbles are joined, each point inside their common boundary becomes exposed
to ionizing photons from both sources. Therefore, the ionizing intensity
inside \ionAR{H}{II} regions rises rapidly, allowing those regions to expand
into high-density gas which had previously recombined fast enough to remain
neutral when the ionizing intensity had been low. Since each bubble
coalescence accelerates the process of reionization, the overlap phase has
the character of a phase transition and is expected to occur rapidly, over
less than a Hubble time at the overlap redshift. By the end of this stage
most regions in the IGM are able to see several unobscured sources, and
therefore the ionizing intensity is much higher than before overlap and it
is also much more homogeneous. An additional ingredient in the rapid
overlap phase results from the fact that hierarchical structure formation
models predict a galaxy formation rate that rises rapidly with time at the
relevant redshift range. This process leads to a state in which the
low-density IGM has been highly ionized and ionizing radiation reaches
everywhere except for gas located inside self-shielded, high-density
clouds. This marks the end of the overlap phase, and this important
landmark is most often referred to as the 'moment of reionization'.

Some neutral gas does, however, remain in high-density structures
which correspond to Lyman Limit systems and damped Ly$\alpha$
systems seen in absorption at lower redshifts. The high-density
regions are gradually ionized as galaxy formation proceeds, and the
mean ionizing intensity also grows with time. The ionizing intensity
continues to grow and to become more uniform as an increasing number
of ionizing sources is visible to every point in the IGM. This
``post-overlap'' phase continues indefinitely, since collapsed objects
retain neutral gas even in the present universe. The IGM does,
however, reach another milestone at $z \sim 1.6$, the breakthrough
redshift (Madau, Haardt, \& Rees 1999). Below this redshift, all
ionizing sources are visible to each other, while above this redshift
absorption by the Ly$\alpha$ forest clouds implies that only
sources in a small redshift range are visible to a typical point in
the IGM.

Semi-analytic models of the pre-overlap stage focus on the evolution
of the \ionAR{H}{II} filling factor $Q_{\rm H\ II}$, i.e., the
fraction of the volume of the universe which is filled by
\ionAR{H}{II} regions. The model of individual \ionAR{H}{II} regions
presented in the previous section can be used to understand the
development of the total filling factor. The overall production of
ionizing photons depends on the collapse fraction $F_{\rm col}$, the
fraction of all the baryons in the universe which are in galaxies,
i.e., the fraction of gas which settles into halos and cools
efficiently inside them. For simplicity, we assume instantaneous
production of photons, i.e., that the timescale for the formation and
evolution of the massive stars in a galaxy is short compared to the
Hubble time at the formation redshift of the galaxy. We also neglect
other complications (such as inhomogeneous clumping and source
clustering) which are discussed below. Under these assumptions, we can
convert equation~(\ref{HIIsoln}), which describes individual
\ionAR{H}{II} regions, to an equation which statistically describes
the transition from a neutral universe to a fully ionized one (see
Barkana \& Loeb 2000c; also Madau et al.\ 1999 and Haiman \& Loeb
1997): \beq Q_{\rm H\ II}(t) =\int_{0}^t \frac{\Ni} {0.76}
\frac{dF_{\rm col}}{dt'}\,e^{F(t',t)} dt'\ , \label{QII} \eeq where
$F(t',t)$ is determined by equation~(\ref{Fgen}).
  
\begin{figure}
\psfig{file=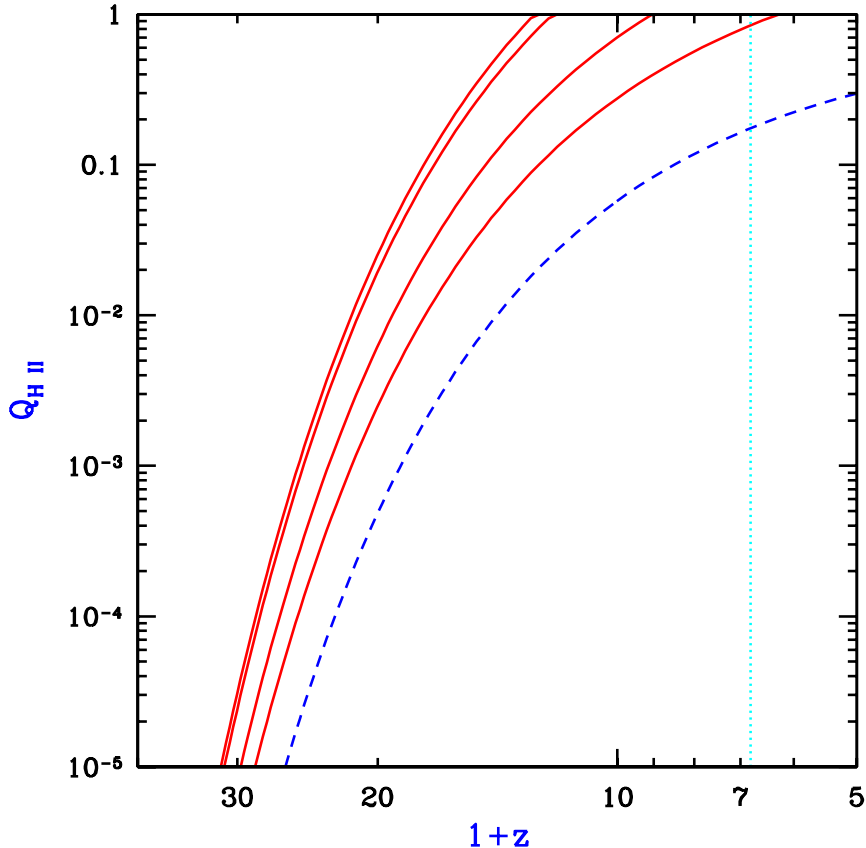,width=4.7in}
%Used FHII.mon
\caption{Semi-analytic calculation of the reionization of the IGM (for
$\Ni=40$), showing the redshift evolution of the filling factor
$Q_{\rm H\ II}$. Solid curves show $Q_{\rm H\ II}$ for a clumping
factor $C=0$ (no recombinations), $C=1$, $C=10$, and $C=30$, in order
from left to right. The dashed curve shows the collapse fraction
$F_{\rm col}$, and the vertical dotted line shows the $z=5.8$
observational lower limit (Fan et al.\ 2000) on the reionization
redshift.}
\label{fig6c}
\end{figure}

A simple estimate of the collapse fraction at high redshift is the
mass fraction [given, e.g., by the model of Press-Schechter (1974)] in
halos above the cooling threshold, which is the minimum mass of halos
in which gas can cool efficiently. Assuming that only atomic cooling
is effective during the redshift range of reionization (Haiman, Rees,
\& Loeb 1997), the minimum mass corresponds roughly to a halo of
virial temperature $T_{\rm vir}=10^4$ K, which corresponds to a total
halo mass of $\sim 8\times 10^7 M_{\sun}$ at $z=10$. With this
prescription we derive (for $\Ni=40$) the reionization history shown
in Figure~\ref{fig6c} for the case of a constant clumping factor
$C$. The solid curves show $Q_{\rm H\ II}$ as a function of redshift
for a clumping factor $C=0$ (no recombinations), $C=1$, $C=10$, and
$C=30$, in order from left to right. Note that if $C \sim 1$ then
recombinations are unimportant, but if $C \ga 10$ then recombinations
significantly delay the reionization redshift (for a fixed
star-formation history). The dashed curve shows the collapse fraction
$F_{\rm col}$ in this model. For comparison, the vertical dotted line
shows the $z=5.8$ observational lower limit (Fan et al.\ 2000) on the
reionization redshift.

%%%%%%%% Figure 
%%%%%%%%
  
Clearly, star-forming galaxies in CDM hierarchical models are capable
of ionizing the universe at $z\sim 6$--15 with reasonable parameter
choices. This has been shown by a number of theoretical, semi-analytic
calculations (Fukugita \& Kawasaki 1994; Shapiro, Giroux, \& Babul
1994; Haiman \& Loeb 1997; Valageas \& Silk 1999; Chiu \& Ostriker
2000; Ciardi et al.\ 2000) as well as numerical simulations (Cen \&
Ostriker 1993; Gnedin \& Ostriker 1997; Gnedin 2000a). Similarly, if a
small fraction ($\la 1\%$) of the gas in each galaxy accretes onto a
central black hole, then the resulting mini-quasars are also able to
reionize the universe, as has also been shown using semi-analytic
models (Fukugita \& Kawasaki 1994; Haiman \& Loeb 1998; Valageas \&
Silk 1999). Note that the prescription whereby a constant fraction of
the galactic mass accretes onto a central black hole is based on local
observations (see \S \ref{sec8.2.2}) which indicate that $z=0$
galaxies harbor central black holes of mass equal to $\sim
0.2$--$0.6\%$ of their bulge mass. Although the bulge constitutes only
a fraction of the total baryonic mass of each galaxy, the higher
gas-to-stellar mass ratio in high redshift galaxies, as well as their
high merger rates compared to their low redshift counterparts, suggest
that a fraction of a percent of the total gas mass in high-redshift
galaxies may have contributed to the formation of quasar black holes.

Although many models yield a reionization redshift around 7--12, the
exact value depends on a number of uncertain parameters affecting both
the sources and the overall recombination rate. The source parameters
include the formation efficiency of stars and quasars and the escape
fraction of ionizing photons produced by these sources (\S
\ref{sec6.1}). The formation efficiency of low mass galaxies may also
be reduced by feedback from galactic outflows [e.g., Dekel \& Silk
1986; Mac Low \& Ferrara 1999; see \S 7 of Barkana \& Loeb (2000c) for
a review of this topic]. Even when the clumping is inhomogeneous, the
treatment in equation~(\ref{QII}) is generally valid if $C$ is defined
as in equation~(\ref{clump}), where we take a global volume average of
the square of the density inside ionized regions (since neutral
regions do not contribute to the recombination rate). The resulting
mean clumping factor depends on the density and clustering of sources,
and on the distribution and topology of density fluctuations in the
IGM. Furthermore, the source halos should tend to form in overdense
regions, and the clumping factor is affected by this cross-correlation
between the sources and the IGM density.

Miralda-Escud\'e, Haehnelt, \& Rees (2000) presented a simple model
for the distribution of density fluctuations, and more generally they
discussed the implications of inhomogeneous clumping during
reionization. They noted that as ionized regions grow, they more
easily extend into low-density regions, and they tend to leave behind
high-density concentrations, with these neutral islands being ionized
only at a later stage. They therefore argued that, since at
high-redshift the collapse fraction is low, most of the high-density
regions, which would dominate the clumping factor if they were
ionized, will in fact remain neutral and occupy only a tiny fraction
of the total volume. Thus, the development of reionization through the
end of the overlap phase should occur almost exclusively in the
low-density IGM, and the effective clumping factor during this time
should be $\sim 1$, making recombinations relatively unimportant (see
Figure~\ref{fig6c}). Only in the post-reionization phase,
Miralda-Escud\'e et al.\ (2000) argued, do the high density clouds and
filaments become gradually ionized as the mean ionizing intensity
further increases.

\begin{figure}
\psfig{file=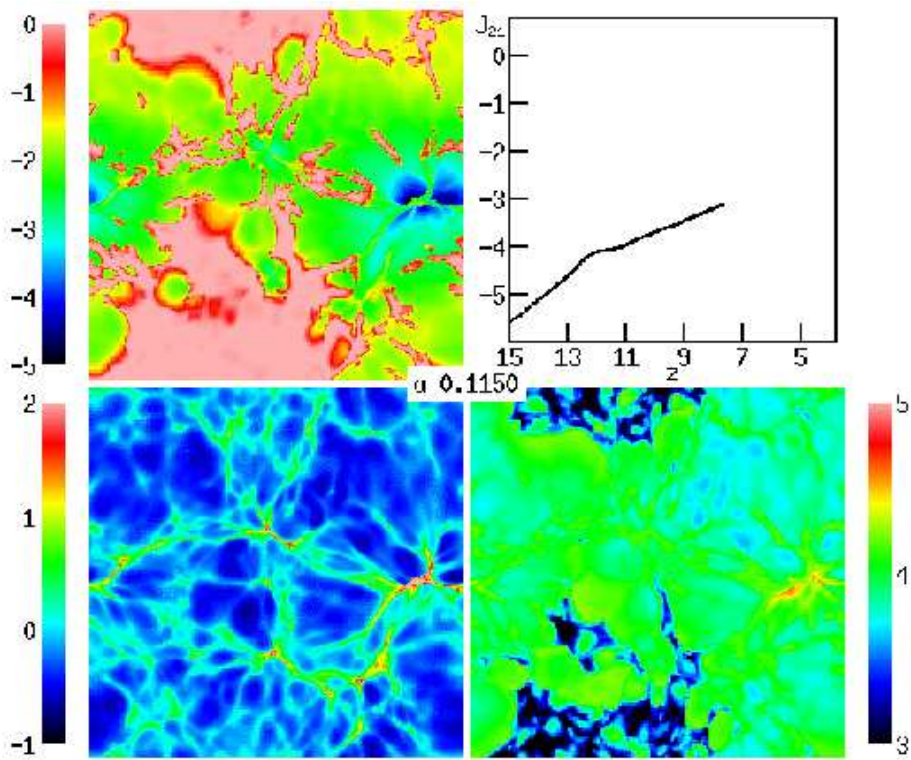,width=5in}
\caption{Visualization at $z=7.7$ of a numerical simulation of
reionization, adopted from Figure~3c of Gnedin (2000a). The panels
display the logarithm of the neutral hydrogen fraction (upper left),
the gas density (lower left), and the gas temperature (lower
right). Also shown is the redshift evolution of the logarithm of the
mean ionizing intensity (upper right). Note the periodic boundary
conditions.}
\label{fig6d}
\end{figure}

\begin{figure}
\psfig{file=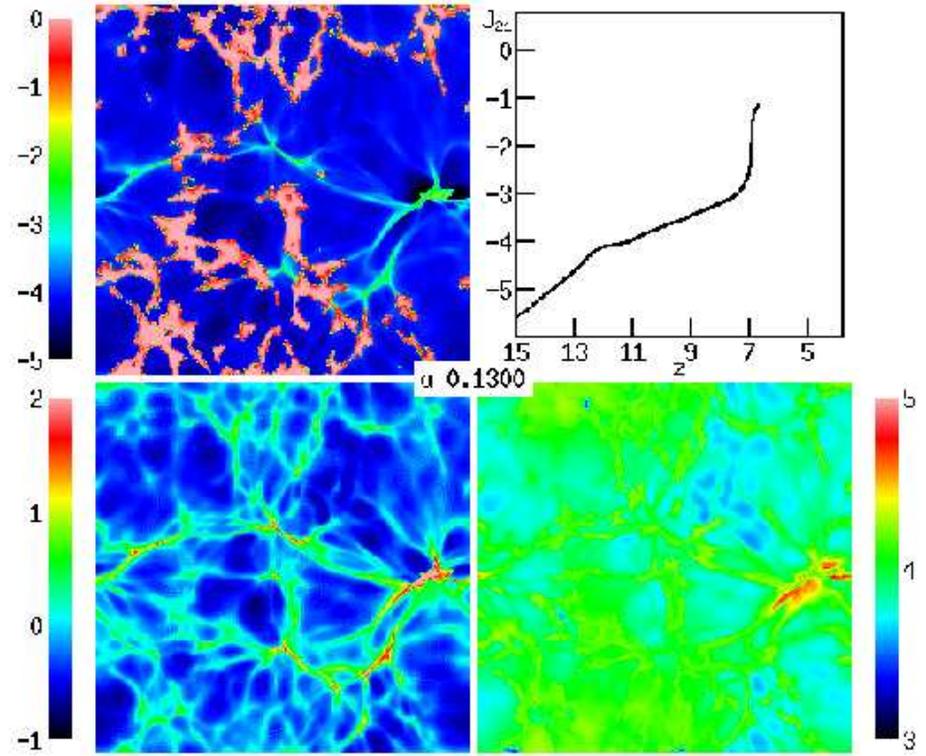,width=5in}
\caption{Visualization at $z=6.7$ of a numerical simulation of
reionization, adopted from Figure~3e of Gnedin (2000a). The panels
display the logarithm of the neutral hydrogen fraction (upper left),
the gas density (lower left), and the gas temperature (lower
right). Also shown is the redshift evolution of the logarithm of the
mean ionizing intensity (upper right). Note the periodic boundary
conditions.}
\label{fig6e}
\end{figure}

The complexity of the process of reionization is illustrated by the
recent numerical simulation by Gnedin (2000a) of stellar reionization
(in $\Lambda$CDM with $\Omm=0.3$). This simulation uses a formulation
of radiative transfer which relies on several rough approximations;
although it does not include the effect of shadowing behind
optically-thick clumps, it does include for each point in the IGM the
effects of an estimated local optical depth around that point, plus a
local optical depth around each ionizing source. This simulation helps
to understand the advantages of the various theoretical approaches,
while pointing to the complications which are not included in the
simple models. Figures~\ref{fig6d} and \ref{fig6e}, taken from
Figure~3 in Gnedin (2000a), show the state of the simulated universe
just before and just after the overlap phase, respectively. They show
a thin (15 $h^{-1}$ comoving kpc) slice through the box, which is 4
$h^{-1}$ Mpc on a side, achieves a spatial resolution of $1 h^{-1}$
kpc, and uses $128^3$ each of dark matter particles and baryonic
particles (with each baryonic particle having a mass of $5\times 10^5
M_{\sun}$).  The figures show the redshift evolution of the mean
ionizing intensity $J_{21}$ (upper right panel), and visually the
logarithm of the neutral hydrogen fraction (upper left panel), the gas
density (lower left panel), and the gas temperature (lower right
panel). Note the obvious features resulting from the periodic boundary
conditions assumed in the simulation. Also note that the intensity
$J_{21}$ is defined as the intensity at the Lyman limit, expressed in
units of $10^{-21}\, \mbox{ erg cm}^{-2} \mbox{ s}^{-1} \mbox{ sr}
^{-1}\mbox{Hz}^{-1}$. For a given source emission, the intensity
inside \ionAR{H}{II} regions depends on absorption and radiative transfer
through the IGM (e.g., Haardt \& Madau 1996; Abel \& Haehnelt 1999)

%%%%%%%% Figure 2
  
%%%%%%%%

%%%%%%%% Figure6.5
  
%%%%%%%%

Figure~\ref{fig6d} shows the two-phase IGM at $z=7.7$, with ionized
bubbles emanating from one main concentration of sources (located at
the right edge of the image, vertically near the center; note the
periodic boundary conditions). The bubbles are shown expanding into
low density regions and beginning to overlap at the center of the
image. The topology of ionized regions is clearly complex: While the
ionized regions are analogous to islands in an ocean of neutral
hydrogen, the islands themselves contain small lakes of dense neutral
gas. One aspect which has not been included in theoretical models of
clumping is clear from the figure. The sources themselves are located
in the highest density regions (these being the sites where the
earliest galaxies form) and must therefore ionize the gas in their
immediate vicinity before the radiation can escape into the low
density IGM. For this reason, the effective clumping factor is of
order 100 in the simulation and also, by the overlap redshift, roughly
ten ionizing photons have been produced per baryon. Figure~\ref{fig6e}
shows that by $z=6.7$ the low density regions have all become highly
ionized along with a rapid increase in the ionizing intensity. The
only neutral islands left are the highest density regions (compare the
two panels on the left). However, we emphasize that the quantitative
results of this simulation must be considered preliminary, since the
effects of increased resolution and a more accurate treatment of
radiative transfer are yet to be explored. Methods are being developed
for incorporating a more complete treatment of radiative transfer into
three dimensional cosmological simulations (e.g., Abel, Norman, \&
Madau 1999; Razoumov \& Scott 1999).

Gnedin, Ferrara, \& Zweibel (2000) investigated an additional effect
of reionization. They showed that the Biermann battery in cosmological
ionization fronts inevitably generates coherent magnetic fields of an
amplitude $\sim 10^{-19}$ Gauss. These fields form as a result of the
breakout of the ionization fronts from galaxies and their propagation
through the \ionAR{H}{I} filaments in the IGM. Although the fields are
too small to directly affect galaxy formation, they could be the seeds
for the magnetic fields observed in galaxies and X-ray clusters today.

If quasars contribute substantially to the ionizing intensity during
reionization then several aspects of reionization are modified
compared to the case of pure stellar reionization. First, the ionizing
radiation emanates from a single, bright point-source inside each host
galaxy, and can establish an escape route (\ionAR{H}{II} funnel) more
easily than in the case of stars which are smoothly distributed
throughout the galaxy (\S \ref{sec6.1}). Second, the hard photons
produced by a quasar penetrate deeper into the surrounding neutral
gas, yielding a thicker ionization front.  Finally, the quasar X-rays
catalyze the formation of $H_2$ molecules and allow stars to keep
forming in very small halos (Haiman, Abel, \& Rees 1999).

Oh (2000) showed that star-forming regions may also produce
significant X-rays at high redshift. The emission is due to inverse
Compton scattering of CMB photons off relativistic electrons in the
ejecta, as well as thermal emission by the hot supernova remnant. The
spectrum expected from this process is even harder than for typical
quasars, and the hard photons photoionize the IGM efficiently by
repeated secondary ionizations. The radiation, characterized by
roughly equal energy per logarithmic frequency interval, would produce
a uniform ionizing intensity and lead to gradual ionization and
heating of the entire IGM. Thus, if this source of emission is indeed
effective at high redshift, it may have a crucial impact in changing
the topology of reionization. Even if stars dominate the emission, the
hardness of the ionizing spectrum depends on the initial mass
function. At high redshift it may be biased toward massive,
efficiently ionizing stars (e.g., Abel, Bryan, \& Norman 2000; Bromm,
Coppi, \& Larson 1999), but this remains very much uncertain.

Madau et al.\ (1999) and Miralda-Escud\'e et al.\ (2000) have studied
the possible ionizing sources which brought about reionization by
extrapolating from the observed populations of galaxies and quasars to
higher redshift. The general conclusion is that a high-redshift source
population similar to the one observed at $z=3$--4 would produce
roughly the needed ionizing intensity for reionization. A precise
conclusion, however, remains elusive because of the same kinds of
uncertainties as those found in the models based on CDM: The typical
escape fraction, and the faint end of the luminosity function, are
both not well determined even at $z=3$--4, and in addition the
clumping factor at high redshift must be known in order to determine
the importance of recombinations. Future direct observations of the
source population at redshifts approaching reionization may help
resolve some of these questions.

%<>
\subsection{Helium Reionization}
\label{sec6.3.2}

The sources that reionized hydrogen very likely caused the single
reionization of helium from \ionAR{He}{I} to \ionAR{He}{II}. Neutral helium is
ionized by photons of 24.6 eV or higher energy, and its recombination rate
is roughly equal to that of hydrogen. On the other hand, the ionization
threshold of \ionAR{He}{II} is 54.4 eV, and fully ionized helium recombines
$\ga 5$ times faster than hydrogen. This means that for both quasars and
galaxies, the reionization of \ionAR{He}{II} should occur later than the
reionization of hydrogen, even though the number of helium atoms is smaller
than hydrogen by a factor of 13.  The lower redshift of \ionAR{He}{II}
reionization makes it more accessible to observations and allows it to
serve in some ways as an observational preview of hydrogen reionization.

\begin{figure}
\psfig{file=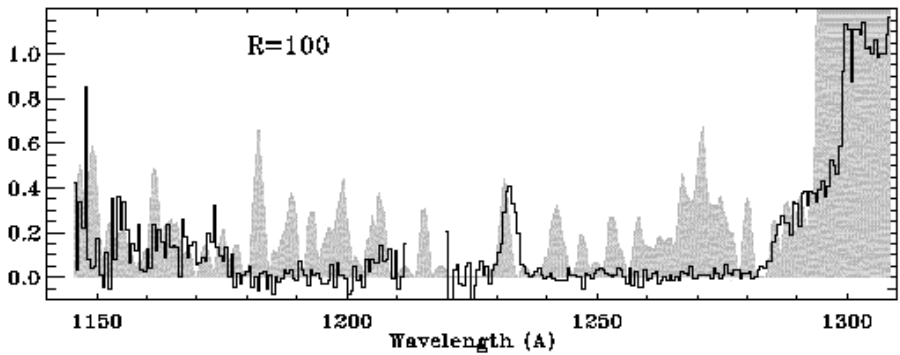,width=5in}
\caption{Ly$\alpha$ absorption by intergalactic \ionAR{He}{II}. This
spectrum of the $z=3.286$ quasar $Q\ 0302-003$, adopted from Figure~4
of Heap et al.\ (2000), was obtained using the Space Telescope Imaging
Spectrograph. The observed spectrum (solid line) is compared to a
simulated spectrum (gray shading) based on the \ionAR{H}{I} Ly$\alpha$
forest observed in the same quasar. In deriving the simulated
spectrum, Heap et al.\ assumed a ratio of \ionAR{He}{II} to \ionAR{H}{I}
column densities of 100, and pure turbulent line broadening.}
\label{VI42fig1}
\end{figure}

The Ly$\alpha$ absorption by intergalactic \ionAR{He}{II} (at wavelength
$304$\AA) has been observed in four quasars at redshifts $2.4 < z <
3.2$ (Jakobsen et al.\ 1994; Davidsen et al.\ 1996; Hogan et al.\
1997; Reimers et al.\ 1997; Anderson et al.\ 1999; Heap et al.\
2000). The results are consistent among the different quasars, and we
illustrate them here with one particular spectrum. In
figure~\ref{VI42fig1}, adopted from Figure~4 of Heap et al.\ (2000),
we show a portion of the spectrum of the $z=3.286$ quasar $Q\
0302-003$, obtained with the Space Telescope Imaging Spectrograph
on-board the {\it Hubble Space Telescope}.\, The observed spectrum
(solid line) is compared to a simulated spectrum (gray shading) based
on the \ionAR{H}{I} Ly$\alpha$ forest observed in the same quasar. In
deriving the simulated spectrum, Heap et al.\ assumed a ratio of
\ionAR{He}{II} to \ionAR{H}{I} column densities of 100, and pure turbulent
line broadening. The wavelength range shown in the figure corresponds
to \ionAR{He}{II} Ly$\alpha$ in the redshift range 2.8--3.3.

%%%%%%%% Figure 1 in section VI.3.2 
%%%%%%%%

The observed flux shows a clear break short-ward of the quasar emission
line at an observed $\lambda=1300$ \AA. Relatively near the quasar, at
$\lambda=$1285--1300\AA, a shelf of relatively high transmission is
likely evidence of the 'proximity effect', in which the emission from
the quasar itself creates a highly ionized local region with a reduced
abundance of absorbing ions. In the region at $\lambda= $1240--1280
\AA\ ($z=3.08$--3.21), on the other hand, the very low flux level
implies an average optical depth of $\tau\sim 4.5$--5. Another large
region with average $\tau \sim 4$, a region spanning $\sim 100$
comoving Mpc along the line of site, is evident at $\lambda=
$1180--1210\AA\ ($z =2.88$--2.98). The strong continuous absorption in
these large regions, and the lack of correlation with the observed
\ionAR{H}{I} Ly$\alpha$ forest, is evidence for a \ionAR{He}{II}
Gunn-Peterson absorption trough due to the diffuse IGM. It also
suggests a rather soft UV background with a significant stellar
contribution, i.e., a background that ionizes the diffuse hydrogen
much more thoroughly than \ionAR{He}{II}. Significant emission is
observed in between the two regions of constant high absorption. A
small region around 1216\AA\ is contaminated by geo-coronal
Ly$\alpha$, but the emission at 1230--1235\AA\ apparently corresponds
to a real, distinct gap in the \ionAR{He}{II} abundance, which could be
caused by a local source photo-ionizing a region of radius $\sim 10$
comoving Mpc. The region at $\lambda=$1150--1175\AA\ ($z=2.78$--2.86)
shows a much higher overall transmission level than the regions at
slightly higher redshift. Heap et al.\ measure an average $\tau=1.9$
in this region, and note that the significant correlation of the
observed spectrum with the simulated one suggests that much of the
absorption is due to a \ionAR{He}{II} Ly$\alpha$ forest while the
low-density IGM provides a relatively low opacity in this region. The
authors conclude that the observed data suggest a sharp opacity break
occurring between $z=3.0$ and 2.9, accompanied by a hardening of the UV
ionizing background.  However, even the relatively high opacity at $z
\ga 3$ only requires $\sim 0.1\%$ of helium atoms not to be fully
ionized, in a region at the mean baryon density. Thus, the overlap
phase of full helium reionization may have occurred significantly
earlier, with the ionizing intensity already fairly uniform but still
increasing with time at $z \sim 3$.

The properties of helium reionization have been investigated
numerically by a number of authors. Zheng \& Davidsen (1995) modeled
the \ionAR{He}{II} proximity effect, and a number of authors
(Miralda-Escud\'e et al.\ 1996; Croft et al.\ 1997; Zhang et al.\ 
1998) used numerical simulations to show that the observations
generally agree with cold dark matter models. They also found that
helium absorption particularly tests the properties of under-dense
voids which produce much of the \ionAR{He}{II} opacity but little opacity
in \ionAR{H}{I}. According to the semi-analytic model of inhomogeneous
reionization of Miralda-Escud\'e, Haehnelt, \& Rees (2000; see also \S
\ref{sec6.3.1}), the total emissivity of observed quasars at redshift
3 suffices to completely reionize helium before $z=3$. They find that
the observations at $z\sim 3$ can be reproduced if a population of
low-luminosity sources, perhaps galaxies, has ionized the low-density
IGM up to an overdensity of around 12, with luminous quasars creating
the observed gaps of transmitted flux.

The conclusion that an evolution of the ionization state of helium has
been observed is also strengthened by several indirect lines of
evidence. Songaila \& Cowie (1996) and Songaila (1998) found a rapid
increase in the \ionAR{Si}{4}/\ionAR{C}{4} ratio with decreasing redshift
at $z=3$, for intermediate column density hydrogen Ly$\alpha$
absorption lines. They interpreted this evolution as a sudden
hardening below $z=3$ of the spectrum of the ionizing background.
Boksenberg et al.\ (1998) also found an increase in the
\ionAR{Si}{4}/\ionAR{C}{4} ratio, but their data implied a much more
gradual increase from $z=3.8$ to $z=2.2$. 

The full reionization of helium due to a hard ionizing spectrum should
also heat the IGM to 20,000 K or higher, while the IGM can only reach
$\sim$10,000 K during a reionization of hydrogen alone. This increase
in temperature can serve as an observational probe of helium
reionization, and it should also increase the suppression of dwarf
galaxy formation (\S \ref{sec6.5}). The temperature of the IGM can be
measured by searching for the smallest line-widths among hydrogen
Ly$\alpha$ absorption lines (Schaye et al.\ 1999). In general, bulk
velocity gradients contribute to the line width on top of thermal
velocities, but a lower bound on the width is set by thermal
broadening, and the narrowest lines can be used to measure the
temperature. Several different measurements (Schaye et al.\ 2000;
Ricotti et al.\ 2000; Bryan \& Machacek 2000; McDonald et al.\ 2000)
have found a nearly isothermal IGM at a temperature of $\sim$20,000 K
at $z=3$, higher than expected in ionization equilibrium and
suggestive of photo-heating due to ongoing reionization of
helium. However, the measurement errors remain too large for a firm
conclusion about the redshift evolution of the IGM temperature or its
equation of state.

Clearly, the reionization of helium is already a rich phenomenological
subject. Our knowledge will benefit from measurements of increasing
accuracy, made toward many more lines of sight, and extended to higher
redshift. New ways to probe helium will also be useful. For example,
Miralda-Escud\'e (2000) has suggested that continuum \ionAR{He}{II}
absorption in soft X-rays can be used to determine the \ionAR{He}{II}
fraction along the line of sight, although the measurement requires an
accurate subtraction of the Galactic contribution to the absorption,
based on the Galactic \ionAR{H}{I} column density as determined by 21 cm
maps.

%<>
\subsection{Photo-evaporation of Gaseous Halos After Reionization}
\label{sec6.4}

The end of the reionization phase transition resulted in the emergence
of an intense UV background that filled the universe and heated the
IGM to temperatures of $\sim 1$--$2\times 10^4$K (see the previous
section). After ionizing the rarefied IGM in the voids and filaments
on large scales, the cosmic UV background penetrated the denser
regions associated with the virialized gaseous halos of the first
generation of objects. A major fraction of the collapsed gas had been
incorporated by that time into halos with a virial temperature $\la
10^4$K, where the lack of atomic cooling prevented the formation of
galactic disks and stars or quasars. Photoionization heating by the
cosmic UV background could then evaporate much of this gas back into
the IGM. The photo-evaporating halos, as well as those halos which did
retain their gas, may have had a number of important consequences just
after reionization as well as at lower redshifts.

In this section we focus on the process by which gas that had already
settled into virialized halos by the time of reionization was
evaporated back into the IGM due to the cosmic UV background. This
process was investigated by Barkana \& Loeb (1999) using semi-analytic
methods and idealized numerical calculations. They first considered an
isolated spherical, centrally-concentrated dark matter halo containing
gas. Since most of the photo-evaporation occurs at the end of overlap,
when the ionizing intensity builds up almost instantaneously, a sudden
illumination by an external ionizing background may be assumed.
Self-shielding of the gas implies that the halo interior sees a
reduced intensity and a harder spectrum, since the outer gas layers
preferentially block photons with energies just above the Lyman limit.
It is useful to parameterize the external radiation field by a
specific intensity per unit frequency, $\nu$, \beq J_{\nu}=10^{-21}\,
J_{21}\, \left(\frac{\nu}{\nu_L}\right)^ {-\alpha}\mbox{ erg
cm}^{-2}\mbox{ s}^{-1}\mbox{ sr} ^{-1}\mbox{Hz}^{-1}\ , \label{Inu}
\eeq where $\nu_L$ is the Lyman limit frequency, and $J_{21}$ is the
intensity at $\nu_L$ expressed in units of $10^{-21}\, \mbox{ erg
cm}^{-2} \mbox{ s}^{-1} \mbox{ sr} ^{-1}\mbox{Hz}^{-1}$. The intensity
is normalized to an expected post--reionization value of around unity
for the ratio of ionizing photon density to the baryon density.
Different power laws can be used to represent either quasar spectra
($\alpha \sim 1.8$) or stellar spectra ($\alpha \sim 5$).

\begin{figure}
%Used avi56.mon (related to avi22b.mon).
%Used avi31.mon
\psfig{file=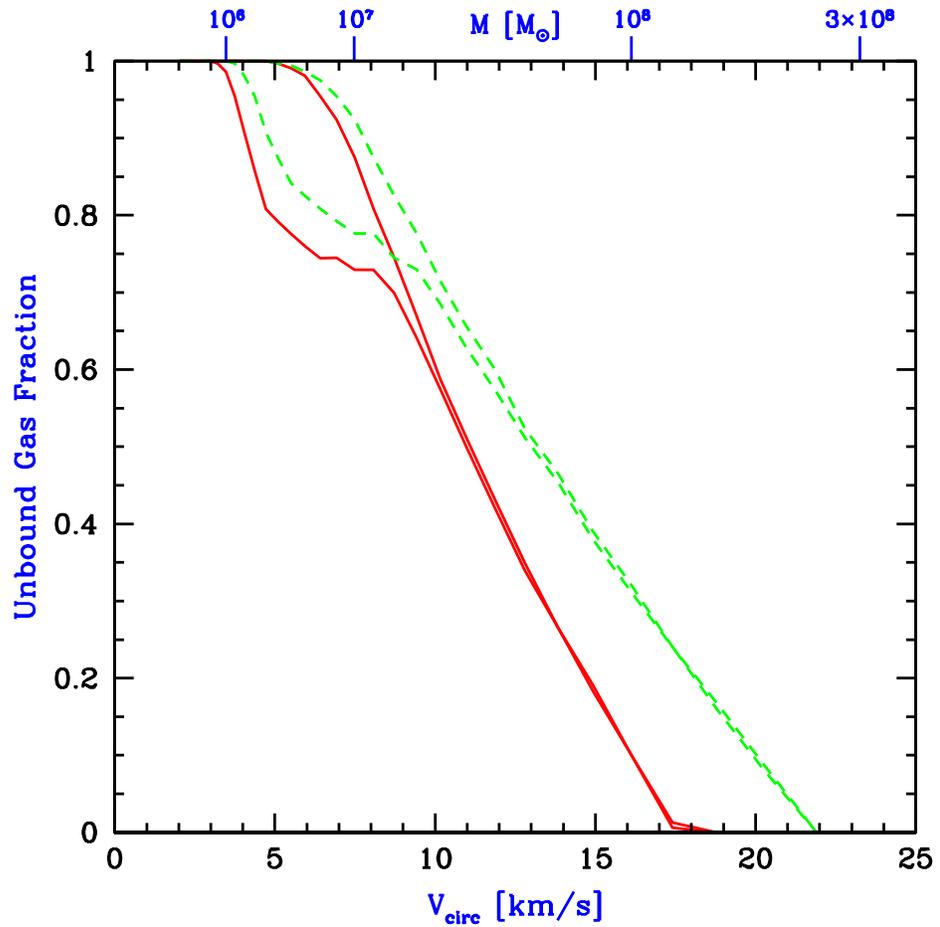,width=5in}
\caption{Effect of photo-evaporation on individual halos. The figure
shows the unbound gas fraction (within the virial radius) versus total
halo velocity dispersion or mass, adopted from Figure~3 of Barkana \&
Loeb (1999). The two pairs of curves correspond to spectral index
$\alpha=5$ (solid) or $\alpha=1.8$ (dashed), in each case at $z=8$. In
each pair, assuming an optically-thin halo leads to the upper curve,
while the lower curve shows the result of including radiative transfer
and self shielding.}
\label{fig6.7}
\end{figure}

Once the gas is heated throughout the halo, some fraction of it
acquires a sufficiently high temperature that it becomes unbound. This
gas expands due to the resulting pressure gradient and eventually
evaporates back into the IGM. Figure~\ref{fig6.7} (adopted from
Figure~3 of Barkana \& Loeb 1999) shows the fraction of gas within the
virial radius which becomes unbound after reionization, as a function
of the total halo circular velocity, with halo masses at $z=8$
indicated at the top. The two pairs of curves correspond to spectral
index $\alpha=5$ (solid) or $\alpha=1.8$ (dashed). In each pair, a
calculation which assumes an optically-thin halo leads to the upper
curve, but including radiative transfer and self-shielding modifies
the result to the one shown by the lower curve. In each case
self-shielding lowers the unbound fraction, but it mostly affects only
a neutral core containing $\sim 30\%$ of the gas. Since high energy
photons above the Lyman limit penetrate deep into the halo and heat
the gas efficiently, a flattening of the spectral slope from
$\alpha=5$ to $\alpha=1.8$ raises the unbound gas fraction. The
characteristic circular velocity where most of the gas is lost is
$\sim 10$--$15~{\rm km~s^{-1}}$ (roughly independent of redshift), but
clearly the effect of photo-evaporation is gradual, going from total
gas removal down to no effect over a range of a factor of $\sim 100$
in halo mass.

%%%%%%%% Figure
%%%%%%%%

%%%%%%%% Figure
\begin{figure}
%Used avi56.mon (related to avi22b.mon).
%Used avi31.mon
\psfig{file=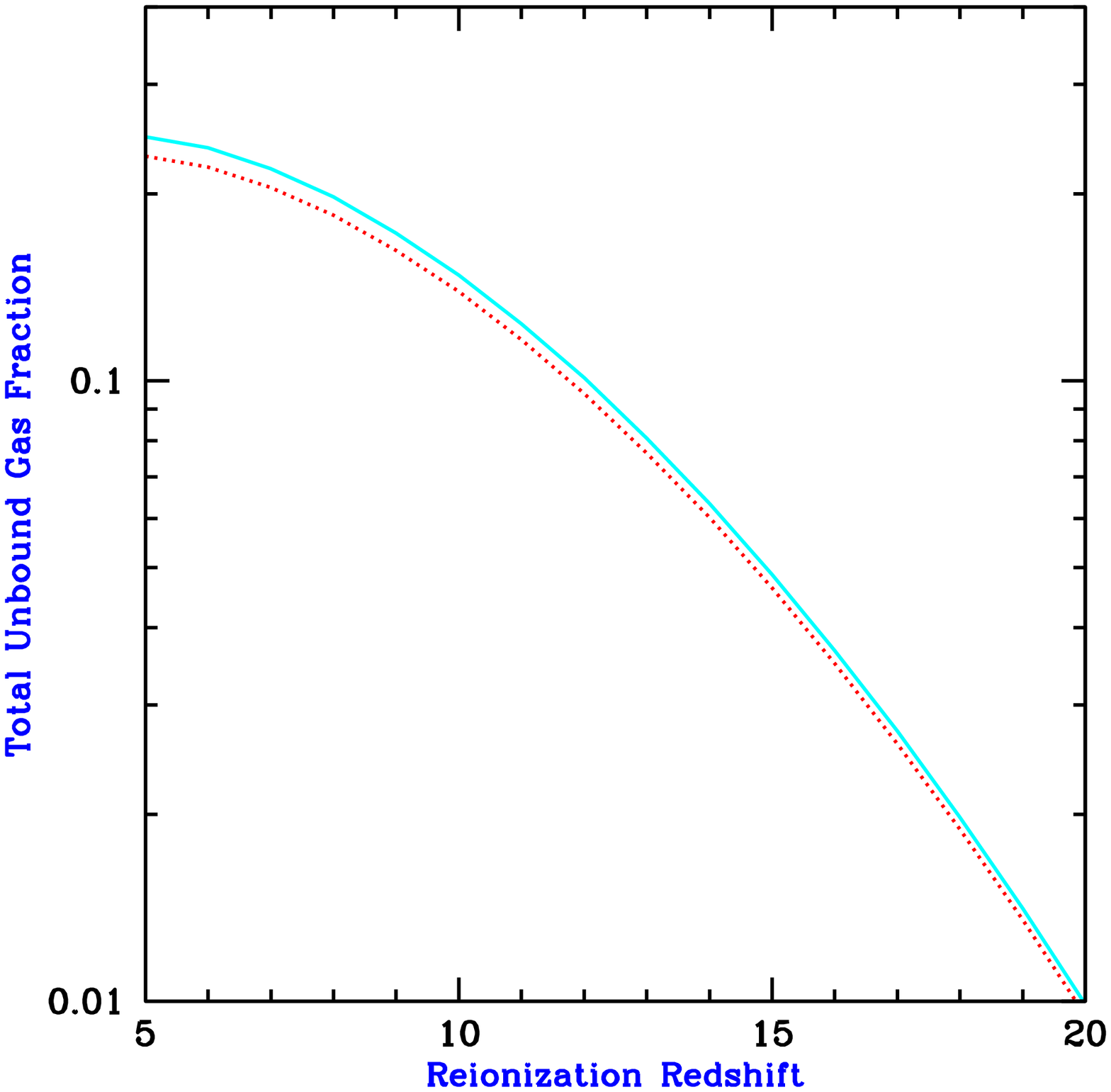,width=5in}
\caption{Effect of photo-evaporation on the overall halo
population. The figure shows the total fraction of gas in the universe
which evaporates from halos at reionization, versus the reionization
redshift, adopted from Figure~7 of Barkana \& Loeb (1999). The solid
line assumes a spectral index $\alpha=1.8$, and the dotted line
assumes $\alpha=5$.}
\label{fig6.7b}
\end{figure}
%%%%%%%%

Given the values of the unbound gas fraction in halos of different
masses, the mass function of Press \& Schechter (1974) can be used to
calculate the total fraction of the IGM which goes through the process
of accreting onto a halo and then being recycled into the IGM at
reionization. The low-mass cutoff in this sum over halos is given by
the lowest mass halo in which gas has assembled by the reionization
redshift. This mass can be estimated by the linear Jeans mass (e.g.,
\S 6 of Peebles 1993), which is $\sim 10^4 M_\odot$ at $z=15$. The
Jeans mass is sufficiently accurate since at $z\sim 10$--20 it agrees
well with the values found in the numerical spherical collapse
calculations of Haiman, Thoul, \& Loeb (1996).

Figure~\ref{fig6.7b} (adopted from Figure~7 of Barkana \& Loeb 1999)
shows the total fraction of gas in the universe which evaporates from
halos at reionization, versus the reionization redshift. The solid
line assumes a spectral index $\alpha=1.8$, and the dotted line
assumes $\alpha=5$, showing that the result is insensitive to the
spectrum. Even at high redshift, the amount of gas which participates
in photo-evaporation is significant, which suggests a number of
possible implications as discussed below. The gas fraction shown in
the figure represents most ($\sim 60$--$80\%$ depending on the
redshift) of the collapsed fraction before reionization, although some
gas does remain in more massive halos.

The photo-evaporation of gas out of large numbers of halos may have
interesting implications. For instance, the resulting $\sim 20~{\rm
km~s^{-1}}$ outflows induce small-scale fluctuations in peculiar
velocity and temperature. These outflows are usually well below the
resolution limit of most numerical simulations, but some outflows were
resolved in the simulation of Bryan et al.\ (1998). The evaporating
halos may consume a significant number of ionizing photons in the
post-overlap stage of reionization (e.g., Haiman, Abel, \& Madau
2000), but a definitive determination requires detailed simulations
which include the three-dimensional geometry of source halos and sink
halos.

Although gas is quickly expelled out of the smallest halos,
photo-evaporation occurs more gradually in larger halos which retain
some of their gas. These surviving halos initially expand but they
continue to accrete dark matter and to merge with other halos. These
evaporating gas halos could contribute to the high column density end
of the Ly$\alpha$ forest (Bond, Szalay, \& Silk 1988). Abel \& Mo
(1998) suggested that, based on the expected number of surviving
halos, a large fraction of the Lyman limit systems at $z\sim 3$ may
correspond to mini-halos that survived reionization. Surviving halos
may even have identifiable remnants in the present universe (e.g.,
Bullock et al.\ 2000; Gnedin 2000c; see \S 9.3 of Barkana \& Loeb
2000c for a review of this topic). These ideas thus offer the
possibility that a population of halos which originally formed prior
to reionization may correspond almost directly to several populations
that are observed much later in the history of the universe. However,
the detailed dynamics of photo-evaporating halos are complex, and
detailed simulations are required to confirm these ideas.
Photo-evaporation of a gas cloud has been followed in a two
dimensional simulation with radiative transfer, by Shapiro \& Raga
(2000). They found that an evaporating halo would indeed appear in
absorption as a damped Ly$\alpha$ system initially, and as a weaker
absorption system subsequently. Future simulations will clarify the
contribution to quasar absorption lines of the entire population of
photo-evaporating halos.

\subsection{Suppression of the Formation of Low Mass Galaxies}
\label{sec6.5}

At the end of overlap, the cosmic ionizing background increased
sharply, and the IGM was heated by the ionizing radiation to a
temperature $\ga 10^4$ K. Due to the substantial increase in the IGM
temperature, the intergalactic Jeans mass increased dramatically,
changing the minimum mass of forming galaxies (Rees 1986; Efstathiou
1992; Gnedin \& Ostriker 1997; Miralda-Escud\'e \& Rees 1998).

Gas infall depends sensitively on the Jeans mass. When a halo more
massive than the Jeans mass begins to form, the gravity of its dark
matter overcomes the gas pressure. Even in halos below the Jeans mass,
although the gas is initially held up by pressure, once the dark
matter collapses its increased gravity pulls in some gas (Haiman,
Thoul, \& Loeb 1996). Thus, the Jeans mass is generally higher than
the actual limiting mass for accretion. Before reionization, the IGM
is cold and neutral, and the Jeans mass plays a secondary role in
limiting galaxy formation compared to cooling. After reionization, the
Jeans mass is increased by several orders of magnitude due to the
photoionization heating of the IGM, and hence begins to play a
dominant role in limiting the formation of stars. Gas infall in a
reionized and heated universe has been investigated in a number of
numerical simulations. Thoul \& Weinberg (1996) inferred, based on a
spherically-symmetric collapse simulation, a reduction of $\sim 50\%$
in the collapsed gas mass due to heating, for a halo of circular
velocity $V_c\sim 50\ {\rm km\ s}^{-1}$ at $z=2$, and a complete
suppression of infall below $V_c \sim 30\ {\rm km\ s}^{-1}$. Kitayama
\& Ikeuchi (2000) also performed spherically-symmetric simulations but
included self-shielding of the gas, and found that it lowers the
circular velocity thresholds by $\sim 5\ {\rm km\ s}^{-1}$. Three
dimensional numerical simulations (Quinn, Katz, \& Efstathiou 1996;
Weinberg, Hernquist, \& Katz 1997; Navarro \& Steinmetz 1997) found a
significant suppression of gas infall in even larger halos ($V_c \sim
75\ {\rm km\ s}^{-1}$), but this was mostly due to a suppression of
late infall at $z\la 2$.

When a volume of the IGM is ionized by stars, the gas is heated to a
temperature $T_{\rm IGM}\sim 10^4$ K. If quasars dominate the UV
background at reionization, their harder photon spectrum leads to
$T_{\rm IGM}\sim 2\times 10^4$ K. Including the effects of dark
matter, a given temperature results in a linear Jeans mass (e.g., \S 6
of Peebles 1993) corresponding to a halo circular velocity of $V_J
\sim 80 {\rm km\ s}^{-1}$, for $T_{\rm IGM} =1.5\times 10^4$ K. In
halos with $V_c>V_J$, the gas fraction in infalling gas equals the
universal mean of $\Omega_b/\Omm$, but gas infall is suppressed in
smaller halos. Even for a small dark matter halo, once it collapses to
a virial overdensity of $\Delta_c/\Ommz$ relative to the mean, it can
pull in additional gas. A simple estimate of the limiting circular
velocity, below which halos have essentially no gas infall, is
obtained by substituting the virial overdensity for the mean density
in the definition of the Jeans mass. The resulting estimate is $V_{\rm
lim} \sim 35 {\rm km\ s}^{-1}$, in rough agreement with the numerical
simulations mentioned above.

Although the Jeans mass is closely related to the rate of gas infall
at a given time, it does not directly yield the total gas residing in
halos at a given time. The latter quantity depends on the entire
history of gas accretion onto halos, as well as on the merger
histories of halos, and an accurate description must involve a
time-averaged Jeans mass. Gnedin (2000b) showed that the gas content
of halos in simulations is well fit by an expression which depends on
the filtering mass, a particular time-averaged Jeans mass (Gnedin \&
Hui 1998).

%\clearpage
%<>*************************************************************************

\section{PROPERTIES OF THE EXPECTED SOURCE POPULATION}
\label{sec8}

%<>
\subsection{The Cosmic Star Formation History}
\label{sec8.1}

One of the major goals of the study of galaxy formation is to achieve
an observational determination and a theoretical understanding of the
cosmic star formation history. By now, this history has been sketched
out to a redshift $z\sim 4$ (see, e.g., the compilation of Blain et
al.\ 1999a). This is based on a large number of observations in
different wavebands. These include various
ultraviolet/optical/near-infrared observations (Madau et al.\ 1996;
Gallego et al.\ 1996; Lilly et al.\ 1996; Connolly et al.\ 1997;
Treyer et al.\ 1998; Tresse \& Maddox 1998; Pettini et al.\ 1998a,b;
Cowie, Songaila \& Barger 1999; Gronwall 1999; Glazebrook et al.\ 
1999; Yan et al.\ 1999; Flores et al.\ 1999; Steidel et al.\ 1999). At
the shortest wavelengths, the extinction correction is likely to be
large (a factor of $\sim 5$) and is still highly uncertain. At longer
wavelengths, the star formation history has been reconstructed from
submillimeter observations (Blain et al.\ 1999b; Hughes et al.\ 1998)
and radio observations (Cram 1998). In the submillimeter regime, a
major uncertainty results from the fact that only a minor portion of
the total far infrared emission of galaxies comes out in the observed
bands, and so in order to estimate the star formation rate it is
necessary to assume a spectrum based, e.g., on a model of the dust
emission (see the discussion in Chapman et al.\ 2000). In general,
estimates of the star formation rate (hereafter SFR) apply
locally-calibrated correlations between emission in particular lines
or wavebands and the total SFR. It is often not possible to check
these correlations directly on high-redshift populations, and together
with the other uncertainties (extinction and incompleteness) this
means that current knowledge of the star formation history must be
considered to be a qualitative sketch only.  Despite the relatively
early state of observations, a wealth of new observatories in all
wavelength regions promise to greatly advance the field. In
particular, \NGST\, will be able to detect galaxies and hence determine
the star formation history out to $z\ga 10$.

Hierarchical models have been used in many papers to match observations on
star formation at $z \la 4$ (see, e.g. Baugh et al.\ 1998; Kauffmann \&
Charlot 1998; Somerville \& Primack 1998, and references therein). In
this section we focus on theoretical predictions for the cosmic star
formation rate at higher redshifts. The reheating of the IGM during
reionization suppressed star formation inside the smallest halos (\S
\ref{sec6.5}). Reionization is therefore predicted to cause a drop in
the cosmic SFR. This drop is accompanied by a dramatic fall in the
number counts of faint galaxies. Barkana \& Loeb (2000b) argued that a
detection of this fall in the faint luminosity function could be used
to identify the reionization redshift observationally.

A model for the SFR can be constructed based on the extended
Press-Schechter theory (e.g., Lacey \& Cole 1993). Once a dark matter
halo has collapsed and virialized, the two requirements for forming
new stars are gas infall and cooling. We assume that by the time of
reionization, photo-dissociation of molecular hydrogen (Haiman, Rees,
\& Loeb 1997) has left only atomic transitions as an avenue for
efficient cooling. Before reionization, therefore, galaxies can form
in halos down to a circular velocity of $V_c\sim 17\ {\rm km\
s}^{-1}$, where this limit is set by cooling. On the other hand, when
a volume of the IGM is ionized by stars or quasars, gas infall into
small halos is suppressed (\S \ref{sec6.5}). In order to include a
gradual reionization in the model, we take the simulations of Gnedin
(2000a) as a guide for the redshift interval of reionization.

In general, new star formation in a given galaxy can occur either from
primordial gas or from recycled gas which has already undergone a
previous burst of star formation. Numerical simulations of starbursts
in interacting $z=0$ galaxies (e.g., Mihos \& Hernquist 1994; 1996)
found that a merger triggers significant star formation in a halo even
if it merges with a much less massive partner. Preliminary results
(Somerville 2000, private communication) from simulations of mergers
at $z \sim 3$ find that they remain effective at triggering star
formation even when the initial disks are dominated by gas.
Regardless of the mechanism, we assume that feedback limits the star
formation efficiency, so that only a fraction $\eta$ of the gas is
turned into stars. To derive the luminosity and spectrum resulting
from the stars, we assume an initial mass function which is similar to
the one measured locally (Scalo 1998). We assume a metallicity
$Z=0.001$, and use the stellar population model results of Leitherer
et al.\ (1999) \footnote{Model spectra of star-forming galaxies were
obtained from http://www.stsci.edu/science/starburst99/}. We also
include a Ly$\alpha$ cutoff in the spectrum due to absorption by the
dense Ly$\alpha$ forest. We do not, however, include dust extinction,
which could be significant in some individual galaxies despite the low
mean metallicity expected at high redshift.

Much of the star formation at high redshift is expected to occur in
low mass, faint galaxies, and even \NGST\, may only detect a fraction
of the total SFR. To get a realistic estimate of this fraction, we
include the finite resolution of the instrument as well as a model for
the distribution of disk sizes at each value of halo mass and redshift
(Barkana \& Loeb 2000a). We describe the sensitivity of \NGST\, by
$F_{\nu}^{\rm ps}$, the minimum spectral flux\footnote{Note that
$F_{\nu}^{\rm ps}$ is the total spectral flux of the source, not just
the portion contained within the aperture.}, averaged over wavelengths
0.6--3.5$\mu$m, required to detect a point source. We adopt a value of
$F_{\nu}^{\rm ps}=0.25$ nJy
\footnote{We obtained the flux limit using the \NGST\, calculator at
http://www.ngst.stsci.edu/nms/main/}, assuming a deep 300-hour
integration on an 8-meter \NGST\, and a spectral resolution of
10:1. This resolution should suffice for a $\sim 10\%$ redshift
measurement, based on the Ly$\alpha$ cutoff. 

\begin{figure}
\psfig{file=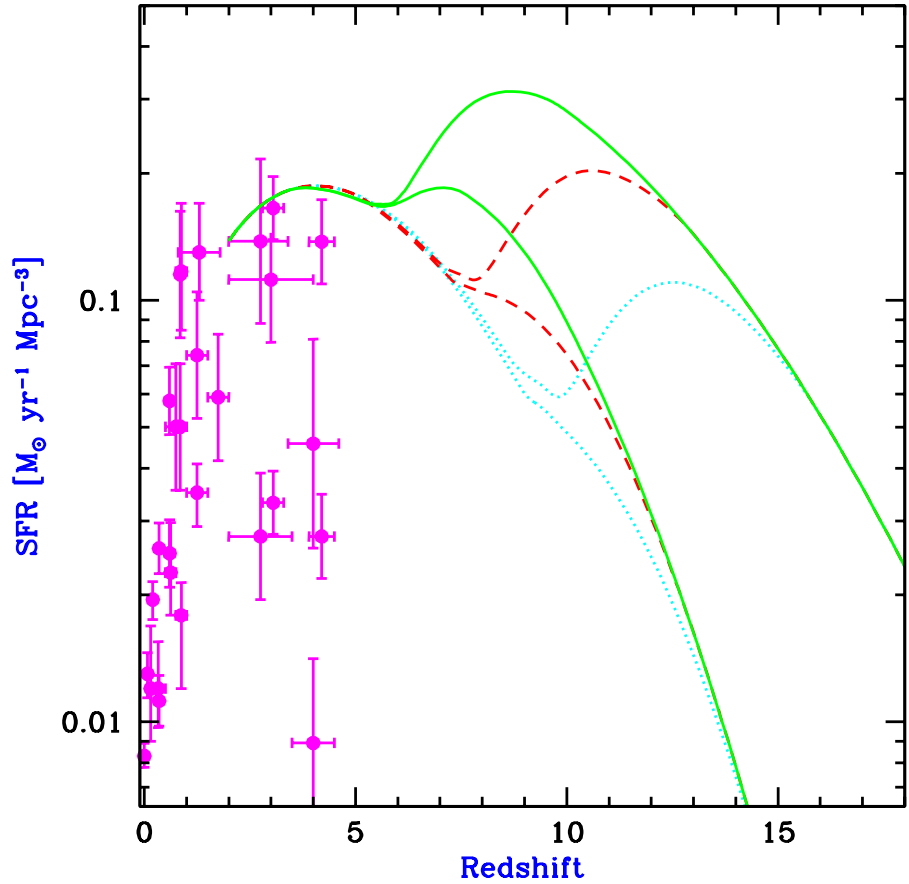,width=5in}
%Used SFDR3.mon, SFR5.f
\caption{Redshift evolution of the SFR (in $M_{\sun}$ per year per
comoving Mpc$^3$), adopted from Figure~1 of Barkana \& Loeb (2000b)
with slight modifications. Points with error bars are observational
estimates (compiled by Blain et al.\ 1999a). Also shown are model
predictions for a reionization redshift $\zr=6$ (solid curves),
$\zr=8$ (dashed curves), and $\zr=10$ (dotted curves), with a star
formation efficiency $\eta=10\%$. In each pair of curves, the upper
one is the total SFR, and the lower one is the fraction detectable
with \NGST\, at a limiting point source flux of 0.25 nJy.}
\label{fig8n}
\end{figure}

Figure~\ref{fig8n} shows our predictions for the star formation
history of the universe, adopted from Figure~1 of Barkana \& Loeb
(2000b) with slight modifications (in the initial mass function and
the values of the cosmological parameters). Letting $\zr$ denote the
redshift at the end of overlap, we show the SFR for $\zr=6$ (solid
curves), $\zr=8$ (dashed curves), and $\zr=10$ (dotted curves). In
each pair of curves, the upper one is the total SFR, and the lower one
is the fraction detectable with {\it NGST}.\, The curves assume a star
formation efficiency $\eta=10\%$ and an IGM temperature $T_{\rm IGM}=
2\times 10^4$ K. Note that the recycled gas contribution to the {\em
detectable} SFR is dominant at the highest redshifts, since the
brightest, highest mass halos form in mergers of halos which
themselves already contain stars. Thus, even though most stars at $z >
\zr$ form out of primordial, zero-metallicity gas, a majority of stars
in detectable galaxies may form out of the small gas fraction that has
already been enriched by the first generation of stars.

%%%%%%%% Figure 8n
%%%%%%%%

Points with error bars in Figure~\ref{fig8n} are observational
estimates of the cosmic SFR per comoving volume at various redshifts
(as compiled by Blain et al.\ 1999a). We choose $\eta=10\%$ to obtain
a rough agreement between the models and these observations at $z\sim
3$--4. An efficiency of order this value is also suggested by
observations of the metallicity of the Ly$\alpha$ forest at $z=3$
(Haiman \& Loeb 1999b). The SFR curves are roughly proportional to the
value of $\eta$. Note that in reality $\eta$ may depend on the halo
mass, since the effect of supernova feedback may be more pronounced in
small galaxies [see \S 7 of Barkana \& Loeb (2000c) for a review of
this topic]. Figure~\ref{fig8n} shows a sharp rise in the total SFR at
redshifts higher than $\zr$. Although only a fraction of the total SFR
can be detected with {\it NGST},\, the detectable SFR displays a
definite signature of the reionization redshift. However, current
observations at lower redshifts demonstrate the observational
difficulty in measuring the SFR directly. The redshift evolution of
the faint luminosity function provides a clearer, more direct
observational signature. We discuss this topic next.

%<>
\subsection{Galaxy Number Counts}
\label{sec8.2.1}

As shown in the previous section, the cosmic star formation history should
display a signature of the reionization redshift. Much of the increase in
the star formation rate beyond the reionization redshift is due to star
formation occurring in very small, and thus faint, galaxies. This evolution
in the faint luminosity function constitutes the clearest observational
signature of the suppression of star formation after reionization.
  
\begin{figure}
\psfig{file=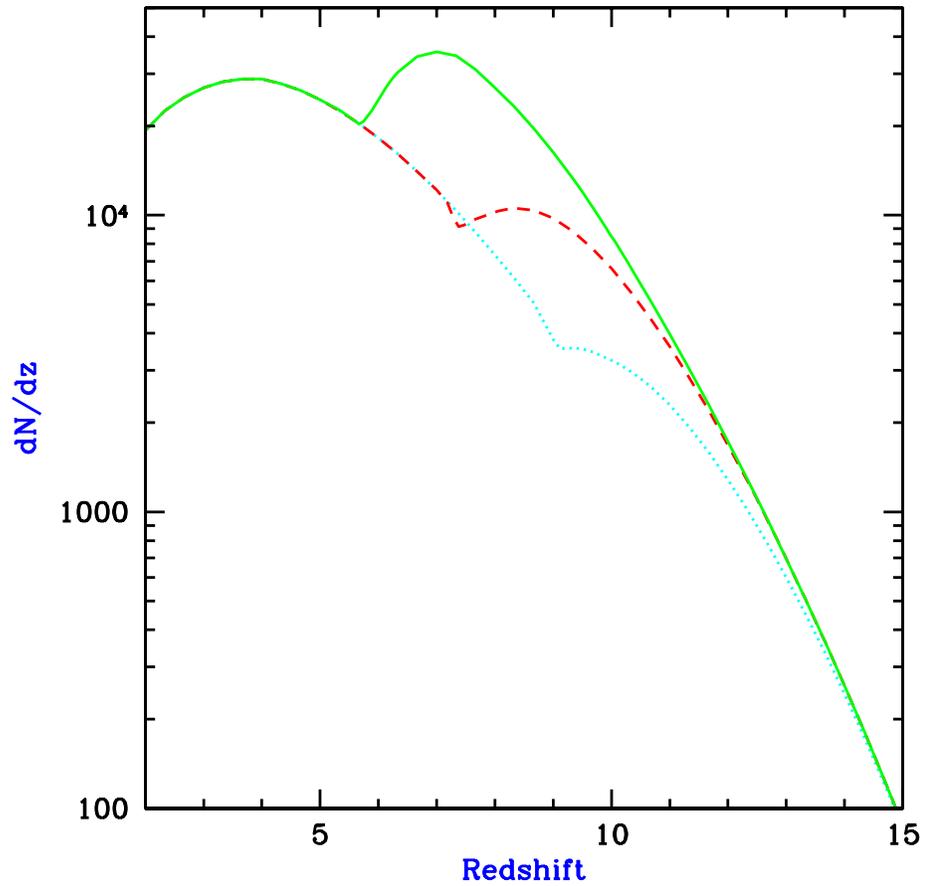,width=5in}
%Used ../SFR/dndz3m.mon
\caption{Predicted redshift distribution of galaxies observed with
{\it NGST},\, adopted and modified from Figure~7 of Barkana \& Loeb
(2000a). The distribution in the $\Lambda$CDM model, with a star
formation efficiency $\eta=10\%$, is shown for a reionization redshift
$\zr=6$ (solid curve), $\zr=8$ (dashed curve), and $\zr=10$ (dotted
curve). The plotted quantity is $dN/dz$, where $N$ is the number of
galaxies per \NGST\, field of view. All curves assume a limiting point
source flux of 0.25 nJy.}
\label{fig8k}
\end{figure}

Figure~\ref{fig8k} shows the predicted redshift distribution in
$\Lambda$CDM of galaxies observed with {\it NGST}.\, The plotted
quantity is $dN/dz$, where $N$ is the number of galaxies per \NGST\,
field of view ($4\arcmin \times 4\arcmin$). The model predictions are
shown for a reionization redshift $\zr=6$ (solid curve), $\zr=8$
(dashed curve), and $\zr=10$ (dotted curve), with a star formation
efficiency $\eta=10\%$. All curves assume a point-source detection
limit of 0.25 nJy. This plot is updated from Figure~7 of Barkana \&
Loeb (2000a) in that redshifts above $\zr$ are included.

%\clearpage

%%%%%%%% Figure 8k
%%%%%%%%
  
\begin{figure}
\psfig{file=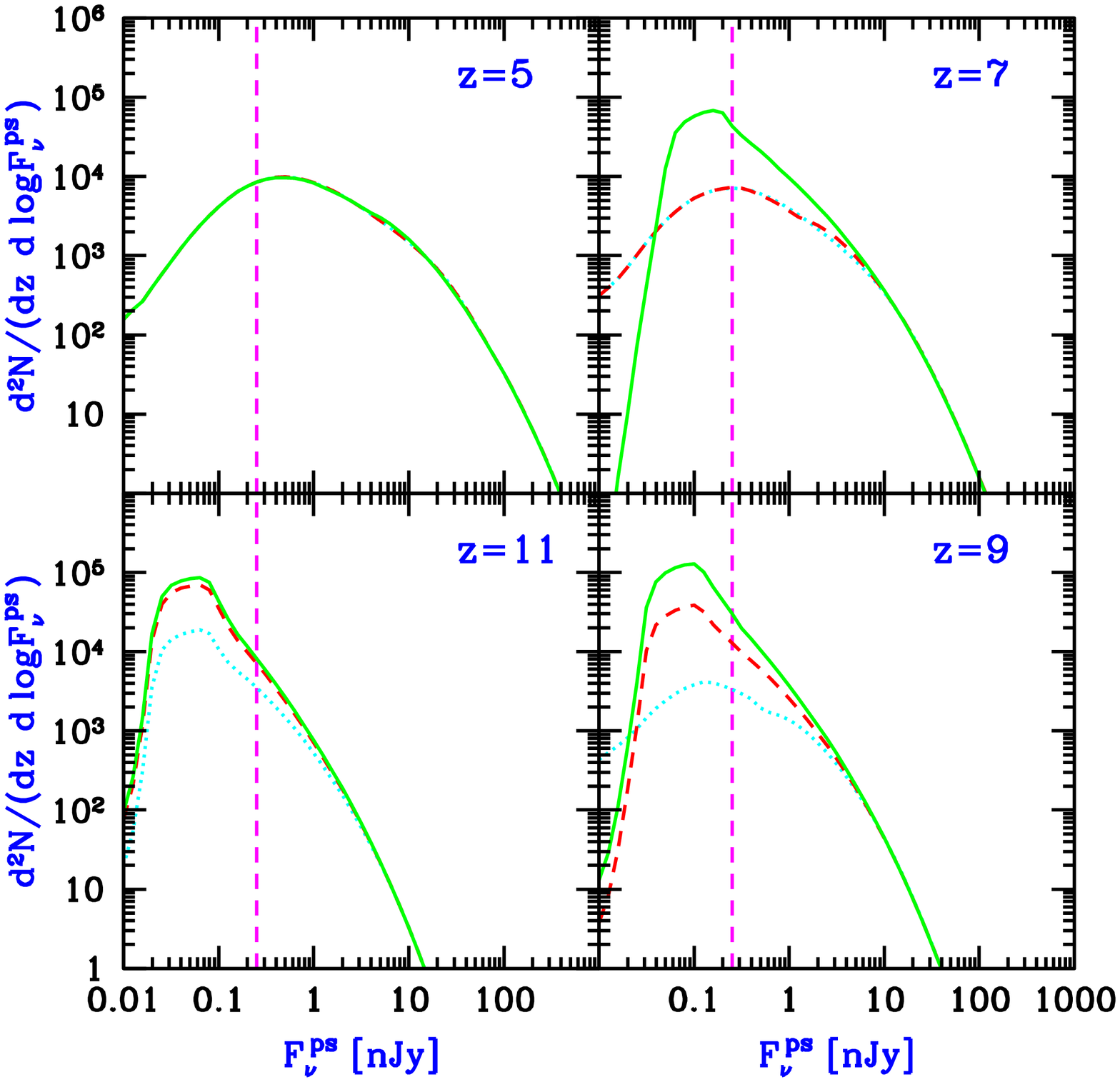,width=4.8in}
%Used ../SFR/dndz4m.mon
\caption{Predicted luminosity function of galaxies at a fixed
redshift, adopted and modified from Figure~2 of Barkana \& Loeb
(2000b). With $\eta=10\%$, the curves show $d^2N/(dz\ d\ln
F_{\nu}^{\rm ps})$, where $N$ is the total number of galaxies in a
single field of view of {\it NGST},\, and $F_{\nu}^{\rm ps}$ is the
limiting point source flux at 0.6--3.5$\mu$m for {\it NGST}.\, The
luminosity function is shown at $z=5$, $z=7$, $z=9$, and $z=11$, with
redshift increasing clockwise starting with the upper left panel. Each
case assumes the $\Lambda$CDM model and a reionization redshift
$\zr=6$ (solid curves), $\zr=8$ (dashed curves), or $\zr=10$ (dotted
curves). The expected \NGST\, detection limit is shown by the vertical
dashed line.}
\label{fig8l}
\end{figure}

Clearly, thousands of galaxies are expected to be found at high
redshift. This will allow a determination of the luminosity function
at many redshift intervals, and thus a measurement of its evolution.
As the redshift is increased, the luminosity function is predicted to
gradually change shape during the overlap era of reionization.
Figure~\ref{fig8l} shows the predicted evolution of the luminosity
function for various values of $\zr$. This Figure is adopted from
Figure~2 of Barkana \& Loeb (2000b) with modifications (in the initial
mass function, the values of the cosmological parameters, and the plot
layout). All curves show $d^2N/(dz\ d\ln F_{\nu}^{\rm ps})$, where $N$
is the total number of galaxies in a single field of view of {\it
NGST},\, and $F_{\nu}^{\rm ps}$ is the limiting point source flux at
0.6--3.5$\mu$m for {\it NGST}.\, Each panel shows the result for a
reionization redshift $\zr=6$ (solid curve), $\zr=8$ (dashed curve),
and $\zr=10$ (dotted curve). Figure~\ref{fig8l} shows the luminosity
function as observed at $z=5$ (upper left panel) and (proceeding
clockwise) at $z=7$, $z=9$, and $z=11$.  Although our model assigns a
fixed luminosity to all halos of a given mass and redshift, in reality
such halos would have some dispersion in their merger histories and
thus in their luminosities. We thus include smoothing in the plotted
luminosity functions. Note the enormous increase in the number density
of faint galaxies in a pre-reionization universe. Observing this
dramatic increase toward high redshift would constitute a clear
detection of reionization and of its major effect on galaxy formation.

%%%%%%%% Figure 8l
  
%<>
\subsection{Quasar Number Counts}
\label{sec8.2.2}

Dynamical studies indicate that massive black holes exist in the
centers of most nearby galaxies (Richstone et al.\ 1998; Kormendy \&
Ho 2000; Kormendy 2000, and references therein). This leads to the
profound conclusion that black hole formation is a generic consequence
of galaxy formation.  The suggestion that massive black holes reside
in galaxies and power quasars dates back to the sixties (Zel'dovich
1964; Salpeter 1964; Lynden Bell 1969).  Efstathiou \& Rees (1988)
pioneered the modeling of quasars in the modern context of galaxy
formation theories. The model was extended by Haehnelt \& Rees (1993)
who added more details concerning the black hole formation efficiency
and lightcurve.  Haiman \& Loeb (1998) and Haehnelt, Natarajan, \&
Rees (1998) extrapolated the model to high redshifts. All of these
discussions used the Press-Schechter theory to describe the abundance
of galaxy halos as a function of mass and redshift. More recently,
Kauffmann \& Haehnelt (2000; also Haehnelt \& Kauffmann 2000) embedded
the description of quasars within semi-analytic modeling of galaxy
formation, which uses the extended Press-Schechter formalism to
describe the merger history of galaxy halos.

In general, the predicted evolution of the luminosity function of
quasars is constrained by the need to match the observed quasar
luminosity function at redshifts $z\la 5$, as well as data from the
Hubble Deep Field (HDF) on faint point--sources.  Prior to
reionization, we may assume that quasars form only in galaxy halos
with a circular velocity $\ga 10~{\rm km~s^{-1}}$ (or equivalently a
virial temperature $\ga 10^4$ K), for which cooling by atomic
transitions is effective. After reionization, quasars form only in
galaxies with a circular velocity $\ga 50\ {\rm km~s^{-1}}$, for which
substantial gas accretion from the warm ($\sim 10^4$ K) IGM is
possible. The limits set by the null detection of quasars in the HDF
are consistent with the number counts of quasars which are implied by
these thresholds (Haiman, Madau, \& Loeb 1999).

For spherical accretion of ionized gas, the bolometric luminosity
emitted by a black hole has a maximum value beyond which radiation
pressure prevents gas accretion. This Eddington luminosity (Eddington
1926) is derived by equating the radiative repulsive force on a free
electron to the gravitational attractive force on an ion in the
plasma. Since both forces scale as $r^{-2}$,
the limiting Eddington luminosity is independent of radius $r$ in the
Newtonian regime, and for gas of primordial composition is given by,
\begin{equation}
L_E=1.45 \times 10^{46} \left({M_{\rm
bh}\over 10^8 M_\odot}\right)~{\rm erg~s^{-1}}\ ,
\end{equation}
where $M_{\rm bh}$ is the black hole mass. Generically, the Eddington
limit applies to within a factor of order unity also to simple
accretion flows in a non-spherical geometry (Frank, King, \& Raine
1992).

The total luminosity of a black hole is related to its mass accretion
rate by the radiative efficiency, $\epsilon$,
\begin{equation}
L=\epsilon \dot{M}_{\rm bh} c^2 .
\end{equation}
For accretion through a thin Keplerian disk onto a Schwarzschild
(non-rotating) black hole, $\epsilon=5.7\%$, while for a maximally rotating
Kerr black hole, $\epsilon=42\%$ (Shapiro \& Teukolsky 1983, p.\ 429).  The
thin disk configuration, for which these high radiative efficiencies are
attainable, exists for $L_{\rm disk}\la 0.5 L_E$ (Laor \& Netzer 1989) .

The accretion time can be defined as
\begin{equation}
\tau={M_{\rm bh}\over {\dot M}_{\rm bh}}= 4\times 10^7~{\rm yr} 
\left(\epsilon \over 0.1\right) \left({L\over L_E}\right)^{-1},
\end{equation}
This time is comparable to the dynamical time inside the central kpc
of a typical galaxy, $t_{\rm dyn}\sim ({\rm 1~kpc}/100~{\rm
km~s^{-1}})=10^7~{\rm yr}$. As long as its fuel supply is not limited
and $\epsilon$ is constant, a black hole radiating at the Eddington
limit will grow its mass exponentially with an $e$--folding time equal
to $\tau$. The fact that $\tau$ is much shorter than the age of the
universe even at high redshift implies that black hole growth is
mainly limited by the feeding rate ${\dot M}_{\rm bh}(t)$, or by the
total fuel reservoir, and not by the Eddington limit.

The ``simplest model'' for quasars involves the following three
assumptions (Haiman \& Loeb 1998):

\noindent
(i) A fixed fraction of the baryons in each ``newly formed'' galaxy
ends up making a central black hole.

\noindent
(ii) Each black hole shines at its maximum (Eddington) luminosity for
a universal amount of time.

\noindent
(iii) All black holes share the same emission spectrum during their
luminous phase (approximated, e.g., by the average quasar spectrum
measured by Elvis et al.\ 1994).

\begin{figure} % fig.1
\psfig{file=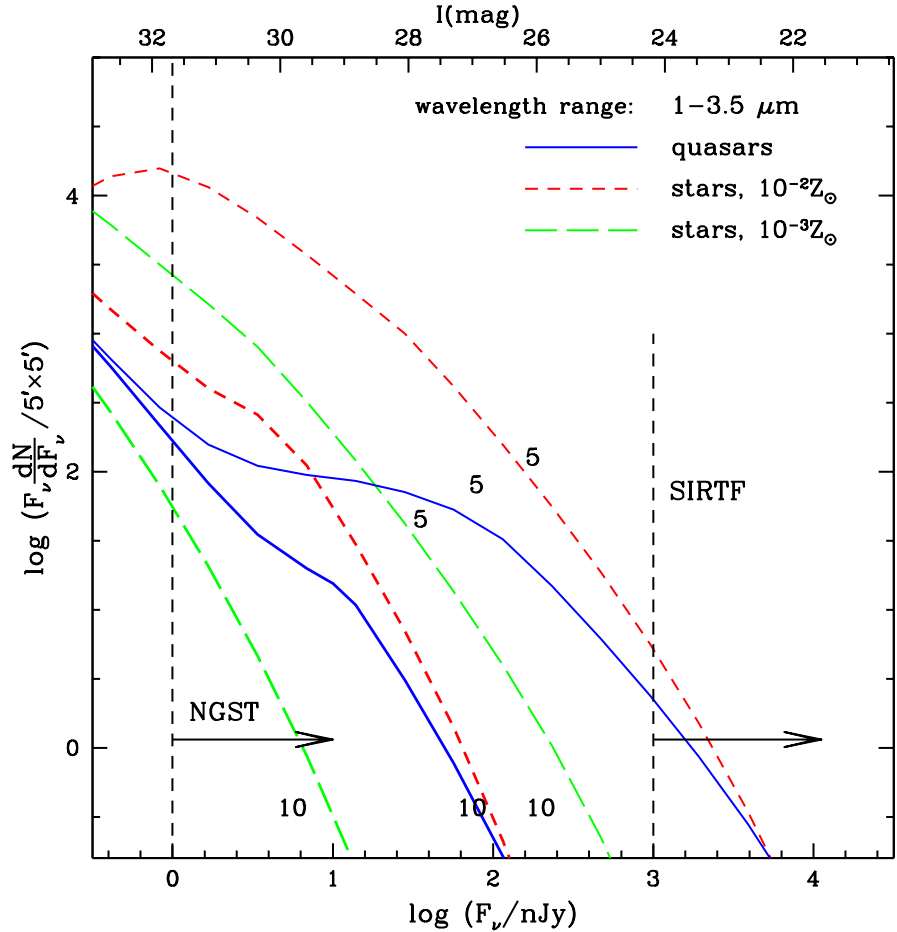,width=5in}
\caption{Infrared number counts of quasars (averaged over the
wavelength interval of $1$--$3.5\mu$m) based on the ``simplest quasar
model'' of Haiman \& Loeb (1999b). The solid curves refer to quasars,
while the long/short dashed curves correspond to star clusters with
low/high normalization for the star formation efficiency.  The curves
labeled ``5'' or ``10'' show the cumulative number of objects with
redshifts above $z=5$ or 10.}
\label{fig8e}
\end{figure}

Note that these assumptions relate only to the most luminous phase of
the black hole accretion process, and they may not be valid during
periods when the radiative efficiency or the mass accretion rate is
very low. At high redshifts the number of ``newly formed'' galaxies
can be estimated based on the time-derivative of the Press-Schechter
mass function, since the collapsed fraction of baryons is small and
most galaxies form out of the unvirialized IGM. Haiman \& Loeb (1998,
1999a) have shown that the above simple prescription provides an
excellent fit to the observed evolution of the luminosity function of
bright quasars between redshifts $2.6<z<4.5$ (see the analytic
description of the existing data in Pei 1995). The observed decline in
the abundance of bright quasars (Schneider, Schmidt, \& Gunn 1991; Pei
1995) results from the deficiency of massive galaxies at high
redshifts. Consequently, the average luminosity of quasars declines
with increasing redshift. The required ratio between the mass of the
black hole and the total baryonic mass inside a halo is $M_{\rm
bh}/M_{\rm gas}=10^{-3.2}\Omm/\Omega_b=5.5\times10^{-3}$, comparable
to the typical value of $\sim 2$--$6\times10^{-3}$ found for the ratio
of black hole mass to spheroid mass in nearby elliptical galaxies
(Magorrian et al.\ 1998; Kormendy 2000). The required lifetime of the
bright phase of quasars is $\sim 10^6$ yr. Figure~\ref{fig8e} shows
the most recent prediction of this model (Haiman \& Loeb 1999a) for
the number counts of high-redshift quasars, taking into account the
above-mentioned thresholds for the circular velocities of galaxies
before and after reionization\footnote{Note that the post-reionization
threshold was not included in the original discussion of Haiman \&
Loeb (1998).}.

We do, however, expect a substantial intrinsic scatter in the ratio
$M_{\rm bh}/M_{\rm gas}$. Observationally, the scatter around the
average value of $\log_{10} (M_{\rm bh}/L)$ is 0.3 (Magorrian et al.\
1998), while the standard deviation in $\log_{10} (M_{\rm bh}/M_{\rm
gas})$ has been found to be $\sigma\sim 0.5$.  Such an intrinsic
scatter would flatten the predicted quasar luminosity function at the
bright end, where the luminosity function is steeply declining.
However, Haiman \& Loeb (1999a) have shown that the flattening
introduced by the scatter can be compensated for through a modest
reduction in the fitted value for the average ratio between the black
hole mass and halo mass by $\sim 50\%$ in the relevant mass range
($10^{8}~{\rm M_\odot}\la M_{\rm bh}\la 10^{10}~{\rm M_\odot}$).

In reality, the relation between the black hole and halo masses may be
more complicated than linear. Models with additional free parameters,
such as a non-linear (mass and redshift dependent) relation between
the black hole and halo mass, can also produce acceptable fits to
the observed quasar luminosity function (Haehnelt et al.\ 1998).  The
nonlinearity in a relation of the type $M_{\rm bh}\propto M_{\rm
halo}^\alpha$ with $\alpha>1$, may be related to the physics of the
formation process of low--luminosity quasars(Haehnelt et al.\ 1998;
Silk \& Rees 1998), and can be tuned so as to reproduce the black hole
reservoir with its scatter in the local universe (Cattaneo, Haehnelt,
\& Rees 1999).  Recently, a tight correlation between the masses of
black holes and the velocity dispersions of the bulges in which they
reside, $\sigma$, was identified in nearby galaxies. Ferrarese \&
Merritt (2000; see also Merritt \& Ferrarese 2000) inferred a
correlation of the type $M_{\rm bh}\propto \sigma^{4.72\pm 0.36}$,
based on a selected sample of a dozen galaxies with reliable $M_{\rm
bh}$ estimates, while Gebhardt et al.\ (2000a,b) have found a somewhat
shallower slope, $M_{\rm bh}\propto \sigma^{3.75(\pm0.3)}$ based on a
significantly larger sample. A non-linear relation of $M_{\rm
bh}\propto \sigma^5 \propto M_{\rm halo}^{5/3}$ has been predicted by
Silk \& Rees (2000) based on feedback considerations, but the observed
relation also follows naturally in the standard semi-analytic models
of galaxy formation (Haehnelt \& Kauffmann 2000).

\begin{figure} % fig8.6
\psfig{file=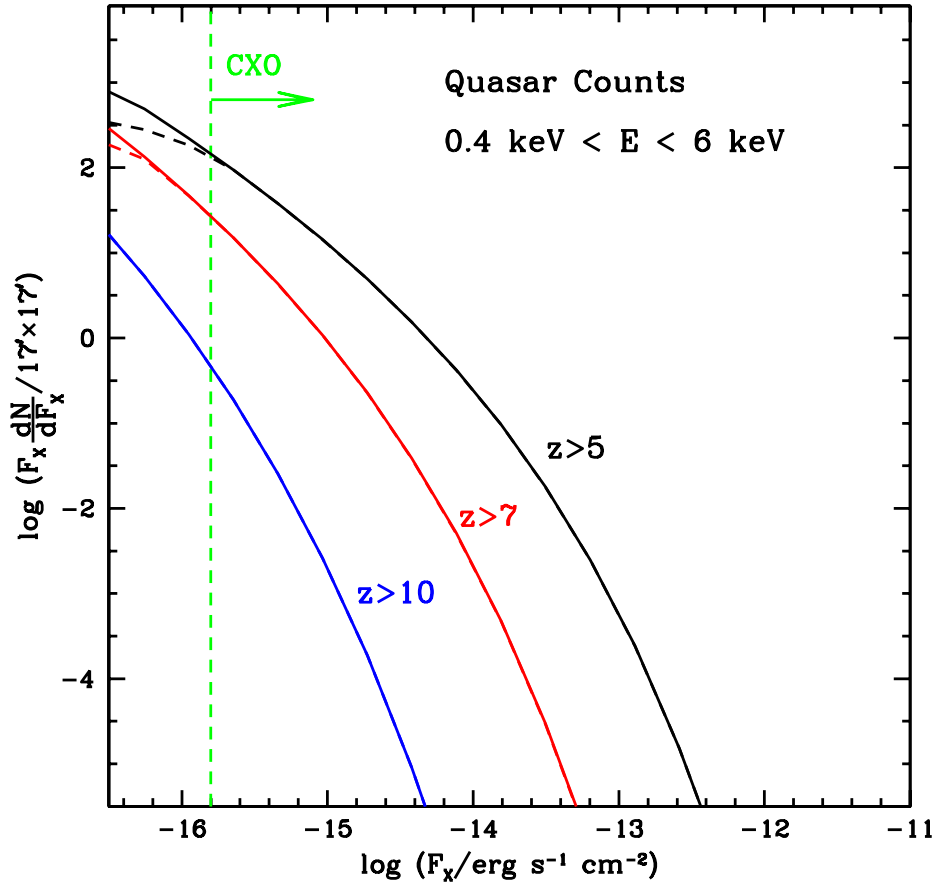,width=5in}
\caption{Total number of quasars with redshift exceeding $z=5$, $z=7$,
and $z=10$ as a function of observed X-ray flux in the {\it CXO}
detection band (from Haiman \& Loeb 1999a). The numbers are normalized
per $17^{\prime}\times 17^{\prime}$ area of the sky. The solid curves
correspond to a cutoff in circular velocity for the host halos of
$v_{\rm circ}\geq 50~{\rm km~s^{-1}}$, the dashed curves to a cutoff
of $v_{\rm circ}\geq 100~{\rm km~s^{-1}}$.  The vertical dashed line
shows the {\it CXO} sensitivity for a 5$\sigma$ detection of a point
source in an integration time of $5\times10^5$ seconds.  }
\label{fig8f}
\end{figure}

Figure~\ref{fig8f} shows the predicted number counts in the ``simplest
model'' described above (Haiman \& Loeb 1999a), normalized to a
$5^{\prime}\times5^{\prime}$ field of view.  Figure~\ref{fig8f} shows
separately the number per logarithmic flux interval of all objects
with redshifts $z>5$ (thin lines), and $z>10$ (thick lines). The
number of detectable sources is high; \NGST\, will be able to probe of
order $100$ quasars at $z>10$, and $\sim200$ quasars at $z>5$ per
$5^{\prime}\times5^{\prime}$ field of view.  The bright--end tail of
the number counts approximately follows a power law, with
$dN/dF_\nu\propto F_\nu^{-2.5}$.  The dashed lines show the
corresponding number counts of ``star--clusters'', assuming that each
halo shines due to a starburst that converts a fraction of 2\%
(long--dashed) or 20\% (short--dashed) of the gas into stars.

\noindent
  
\noindent
  
Similar predictions can be made in the X-ray regime.
Figure~\ref{fig8f} shows the number counts of high-redshift X-ray
quasars in the above ``simplest model''.  This model fits the X-ray
luminosity function of quasars at $z\sim 3.5$ as observed by ROSAT
(Miyaji, Hasinger, \& Schmidt 2000), using the same parameters
necessary to fit the optical data (Pei 1995).  Deep optical or
infrared follow-ups on deep images taken with the Chandra X-ray
satellite (CXO; see, e.g., Mushotzky et al.\ 2000; Barger et al.\
2000; Giaconni et al.\ 2000) will be able to test these predictions in
the relatively near future.

The ``simplest model'' mentioned above predicts that black holes and
stars make comparable contributions to the ionizing background prior
to reionization. Consequently, the reionization of hydrogen and helium
is predicted to occur roughly at the same epoch. A definitive
identification of the \ionAR{He}{II} reionization redshift will provide
another powerful test of this model. Further constraints on the
lifetime of the active phase of quasars may be provided by future
measurements of the clustering properties of quasars (Haehnelt et al.\ 
1998; Martini \& Weinberg 2000; Haiman \& Hui 2000).

%<>
\subsection{High-Redshift Supernovae}
\label{sec8.2.3}

The detection of galaxies and quasars becomes increasingly difficult
at a higher redshift. This results both from the increase in the
luminosity distance and the decrease in the average galaxy mass with
increasing redshift. It therefore becomes advantageous to search for
transient sources, such as supernovae or $\gamma$-ray bursts, as
signposts of high-redshift galaxies (Miralda-Escud\'e \& Rees
1997). Prior to or during the epoch of reionization, such sources are
likely to outshine their host galaxies.

The IGM is observed to be enriched with metals at redshifts $z\la
5$. Identification of \ionAR{C}{IV}, \ionAR{Si}{IV} and \ionAR{O}{VI}
absorption lines which correspond to Ly$\alpha$ absorption lines in
the spectra of high-redshift quasars has revealed that the low-density
IGM has been enriched to a metal abundance (by mass) of $Z_{\rm
IGM}\sim 10^{-2.5 (\pm 0.5)}Z_\odot$, where $Z_\odot=0.019$ is the
solar metallicity (Meyer \& York 1987; Tytler et al.\ 1995; Songaila
\& Cowie 1996; Lu et al 1998; Cowie \& Songaila 1998; Songaila 1997;
Ellison et al.\ 2000). The metal enrichment has been clearly
identified down to \ionAR{H}{I} column densities of $\sim
10^{14.5}~{\rm cm^{-2}}$. The metals detected in the IGM signal the
existence of supernova (SN) explosions at redshifts $z\ga 5$. Since
each SN produces an average of $\sim 1M_\odot$ of heavy elements
(Woosley \& Weaver 1995), the inferred metallicity of the IGM implies
that there should be a supernova at $z\ga 5$ for each $\sim 1.7 \times
10^4M_\odot \times (Z_{\rm IGM}/10^{-2.5} Z_\odot)^{-1}$ of baryons in
the universe. We can therefore estimate the total supernova rate, on
the entire sky, necessary to produce these metals at $z\sim 5$. In a
flat $\Omm=0.3$ cosmology, the total supernova rate across the entire
sky at $z\ga 5$ is estimated to be (Miralda-Escud\'e \& Rees 1997)
roughly one SN per square arcminute per year, if $Z_{\rm
IGM}=10^{-2.5} Z_\odot$.

\begin{figure} % fig8.7
\psfig{file=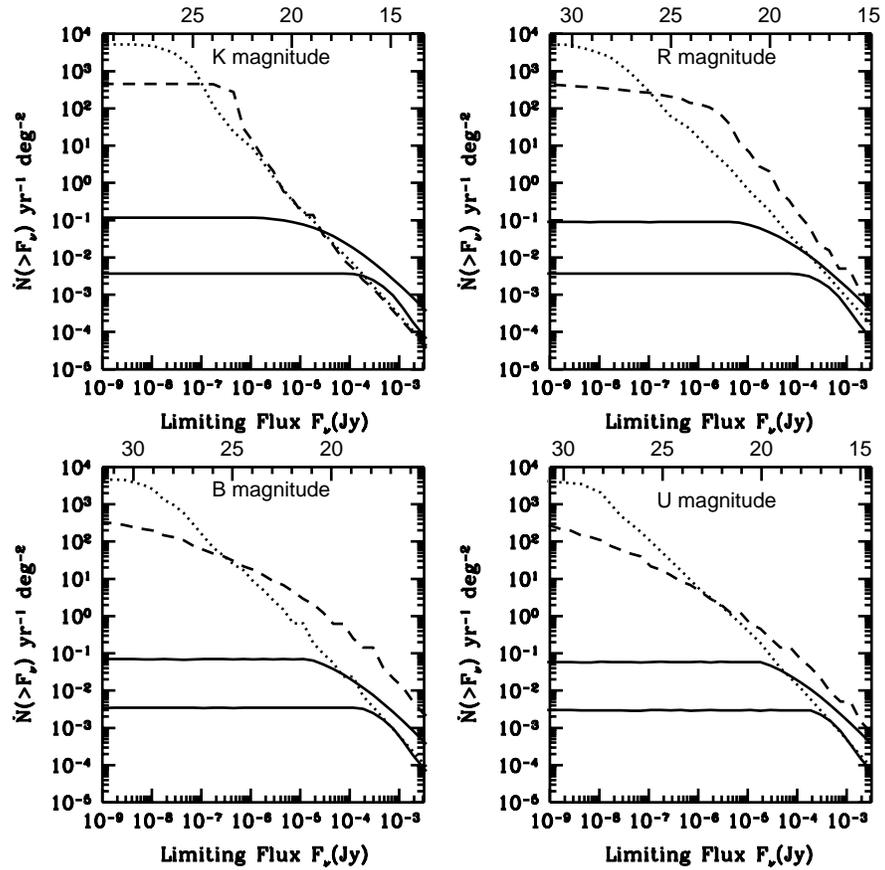,width=4.8in}
%\epsfbox{VIII.2.1_fig3.eps}
\caption{Predicted cumulative rate $\dot{N}(>F_\nu)$ per year per
square degree of supernovae at four wavelengths, corresponding to the
$K$, $R$, $B$, and $U$ bands (from Woods \& Loeb 1998).  The broken
lines refer to different supernova types, namely SNe Ia ({\it dashed
curves}) and SNe II ({\it dotted curves}). For comparison, the solid
curves show estimates for the rates of gamma-ray burst (GRB)
afterglows; the lower solid curve assumes the best-fit rate and
luminosity for GRB sources which trace the star formation history
(Wijers et al.\ 1998), while the upper solid curve assumes the best-fit
values for a non-evolving GRB population.  }
\label{fig8g}
\end{figure}

The actual SN rate at a given observed flux threshold is determined by
the star formation rate per unit comoving volume as a function of
redshift and the initial mass function of stars (e.g., Madau, della
Valle, \& Panagia 1998; Woods \& Loeb 1998; Sullivan et al.\ 2000)
Figure~\ref{fig8g} shows the predicted SN rate as a function of
limiting flux in various bands (Woods \& Loeb 1998), based on the
comoving star formation rate as a function of redshift that was
determined empirically by Madau (1997). The actual star formation rate
may be somewhat higher due to corrections for dust extinction (for a
recent compilation of current data, see Blain \& Natarajan 2000).  The
dashed lines correspond to Type Ia SNe and the dotted lines to Type II
SNe. For comparison, the solid lines indicate two crude estimates for
the rate of $\gamma$--ray burst afterglows, which are discussed in
detail in the next section.

\noindent
  
The above SN rate is appropriate for a threshold experiment, one which
monitors the sky continuously and triggers when the detected flux
exceeds a certain value, and hence identifies the most distant sources
only when they are near their peak flux.  For search strategies which
involve taking a series of ``snapshots'' of a field and looking for
variations in the flux of sources in successive images, one does not
necessarily detect most sources near their peak flux. In this case, we
can estimate the {\it total}\, number of events (i.e., {\it not}\, per
unit time) brighter than $F_\nu$ at a given observed wavelength. A
simple estimate uses the rest-frame duration over which an event will
be brighter than the limiting flux $F_\nu$ at redshift $z$.  This is a
naive estimate of the so-called ``control time''; in practice, the
effective duration over which an event can be observed is shorter,
owing to the image subtraction technique, host galaxy magnitudes, and
a number of other effects which reduce the detection efficiency (Pain
et al.\ 1996). 

Supernovae also produce dust which could process the emission spectrum
of galaxies. Although produced in galaxies, the dust may be expelled
together with the metals out of galaxies by supernova-driven winds.
Loeb \& Haiman (1997) have shown that if each supernova produces $\sim
0.3\, M_\odot$ of Galactic dust, and some of the dust is expelled
together with metals out of the shallow potential wells of the early
dwarf galaxies, then the optical depth for extinction by intergalactic
dust may reach a few tenths at $z\sim 10$ for observed wavelengths of
$\sim 0.5$--$1\,\mu$m [see Todini \& Ferrara (2000) for a detailed
discussion on the production of dust in primordial Type II SNe]. The
opacity in fact peaks in this wavelength band since at $z\sim 10$ it
corresponds to rest-frame UV, where normal dust extinction is most
effective. In these estimates, the amplitude of the opacity is
calibrated based on the observed metallicity of the IGM at $z\la 5$.
The intergalactic dust absorbs the UV background at the reionization
epoch and re-radiates it at longer wavelengths. The flux and spectrum
of the infrared background which is produced at each redshift depends
sensitively on the distribution of dust around the ionizing sources,
since the deviation of the dust temperature from the microwave
background temperature depends on the local flux of UV radiation that
it is exposed to. For reasonable choices of parameters, dust could
lead to a significant spectral distortion of the microwave background
spectrum that could be measured by a future spectral mission, going
beyond the upper limit derived by the COBE satellite (Fixsen et
al. 1996).

The metals produced by supernovae may also yield strong molecular line
emission.  Silk \& Spaans (1997) pointed out that the rotational line
emission of CO by starburst galaxies is enhanced at high redshift due
to the increasing temperature of the cosmic microwave background,
which affects the thermal balance and the level populations of the
atomic and molecular species.  They found that the future Millimeter
Array (MMA) could detect a starburst galaxy with a star formation rate
of $\sim 30\, {\rm M_\odot~yr^{-1}}$ equally well at $z=5$ and $z=30$
because of the increasing cosmic microwave background temperature with
redshift. Line emission may therefore be a more powerful probe of the
first bright galaxies than continuum emission by dust.

%<>
\subsection{High-Redshift Gamma Ray Bursts}
\label{sec8.2.4}

The past decade has seen major observational breakthroughs in the
study of Gamma Ray Burst (GRB) sources. The Burst and Transient Source
Experiment (BATSE) on board the Compton Gamma Ray Observatory (Meegan
et al.\ 1992) showed that the GRB population is distributed
isotropically across the sky, and that there is a deficiency of faint
GRBs relative to a Euclidean distribution.  These were the first
observational clues indicating a cosmological distance scale for GRBs.
The localization of GRBs by X-ray observations with the BeppoSAX
satellite (Costa et al.\ 1997) allowed the detection of afterglow
emission at optical (e.g., van Paradijs et al.\ 1997) and radio (e.g.,
Frail et al.\ 1997) wavelengths up to more than a year following the
events (Fruchter et al.\ 1999; Frail et al.\ 1999).  The afterglow
emission is characterized by a broken power-law spectrum with a peak
frequency that declines with time.  The radiation is well-fitted by a
model consisting of synchrotron emission from a decelerating blast
wave (Blandford \& McKee 1976), created by the GRB explosion in an
ambient medium, with a density comparable to that of the interstellar
medium of galaxies (Waxman 1997; Sari, Piran, \& Narayan 1998; Wijers
\& Galama 1999; M\'esz\'aros 1999; but see also Chevalier \& Li 2000).
The detection of spectral features, such as metal absorption lines in
some optical afterglows (Metzger et al.\ 1997) and emission lines from
host galaxies (Kulkarni et al.\ 2000), allowed an unambiguous
identification of the cosmological distance-scale to these sources.

The nature of the central engine of GRBs is still unknown. Since the
inferred energy release, in cases where it can be securely calibrated
(Freedman \& Waxman 1999; Frail et al.\ 2000), is comparable to that in
a supernova, $\sim 10^{51}~{\rm erg}$, most popular models relate GRBs
to stellar remnants such as neutron stars or black holes (Eichler et
al.\ 1989; Narayan, Paczy\'{n}ski, \& Piran 1992; Paczy\'{n}ski 1991; Usov
1992; Mochkovitch et al.\ 1993; Paczy\'{n}ski 1998; MacFadyen \&
Woosley 1999). Recently it has been claimed that the late evolution of
some rapidly declining optical afterglows shows a component which is
possibly associated with supernova emission (e.g., Bloom et al.\ 1999;
Reichart 1999). If the supernova association is confirmed by detailed
spectra of future afterglows, the GRB phenomenon will be linked to the
terminal evolution of massive stars.

Any association of GRBs with the formation of single compact stars
implies that the GRB rate should trace the star formation history of
the universe (Totani 1997; Sahu et al.\ 1997; Wijers et al.\ 1998; but
see Krumholz, Thorsett \& Harrison 1998).  Owing to their high
brightness, GRB afterglows could in principle be detected out to
exceedingly high redshifts. Just as for quasars, the broad-band
emission of GRB afterglows can be used to probe the absorption
properties of the IGM out to the reionization epoch at redshift $z\sim
10$. Lamb \& Reichart (2000) extrapolated the observed gamma-ray and
afterglow spectra of known GRBs to high redshifts and emphasized the
important role that their detection could play in probing the IGM. In
particular, the broad-band afterglow emission can be used to probe the
ionization and metal-enrichment histories of the intervening
intergalactic medium during the epoch of reionization.

\begin{figure}[htbp] % fig8i
\includegraphics{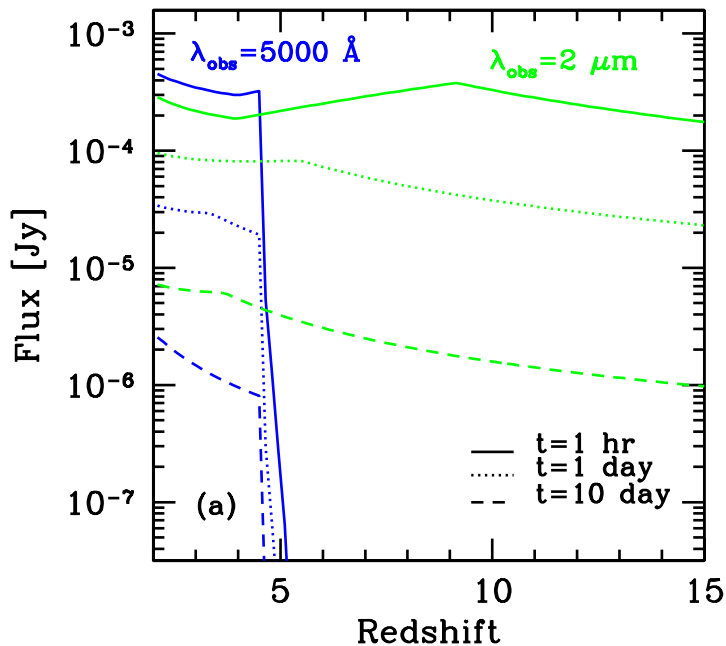}
\vspace{3.3in}
\caption{ Observed flux from a $\gamma$-ray burst afterglow as a
function of redshift (from Ciardi \& Loeb 2000).  The two sets of
curves refer to a photon frequency $\nu=6 \times 10^{14}$ Hz
($\lambda_{obs}=5000$ \AA, thin lines) and $\nu=1.5 \times 10^{14}$ Hz
($\lambda_{obs}=2 \mu$m, thick lines).  Each set shows different
observed times after the GRB trigger; from top to bottom: 1 hour
(solid line), 1 day (dotted) and 10 days (dashed). The sharp
suppression for 5000 \AA ~~at $z\ga 4.5$ is due to IGM absorption.}
\label{fig8i}
\end{figure}

\begin{figure}[htbp] % fig8j
\includegraphics{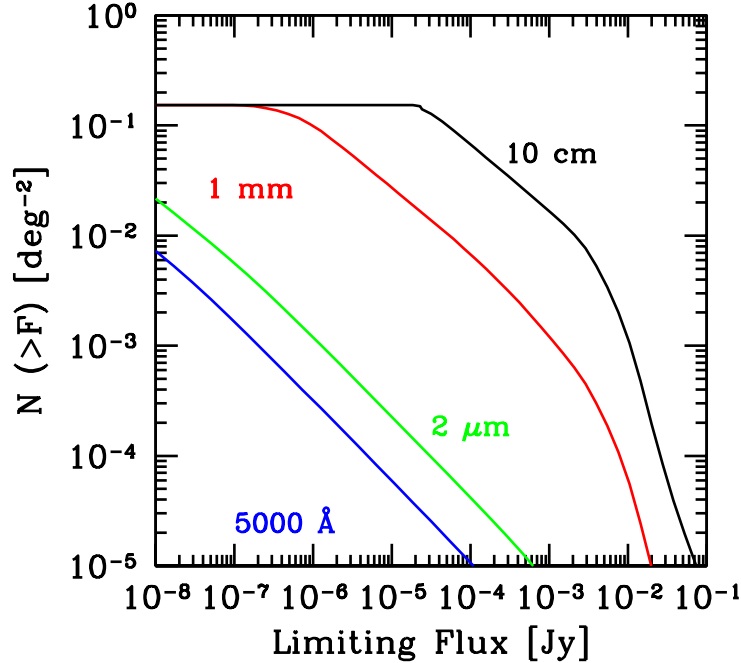}
\vspace{3.4in}
\caption{Predicted number of GRB afterglows per square degree with observed flux
greater than $F$, at several different observed wavelengths (from
Ciardi \& Loeb 2000).  From right to left, the observed wavelength
equals 10 cm, 1 mm, 2 $\mu$m and 5000 \AA.  }
\label{fig8j}
\end{figure}

Ciardi \& Loeb (2000) showed that unlike other sources (such as
galaxies and quasars), which fade rapidly with increasing redshift,
the observed infrared flux from a GRB afterglow at a fixed observed
age is only a weak function of its redshift (Figure~\ref{fig8i}).  A
simple scaling of the long-wavelength spectra and the temporal
evolution of afterglows with redshift implies that at a fixed time-lag
after the GRB in the observer's frame, there is only a mild change in
the {\it observed} flux at infrared or radio wavelengths with
increasing redshift. This results in part from the fact that
afterglows are brighter at earlier times, and that a given observed
time refers to an earlier intrinsic time in the source frame as the
source redshift increases.  The ``apparent brightening'' of GRB
afterglows with redshift could be further enhanced by the expected
increase with redshift of the mean density of the interstellar medium
of galaxies (Wood \& Loeb 2000). Figure~\ref{fig8j} shows the expected
number counts of GRB afterglows, assuming that the GRB rate is
proportional to the star formation rate and that the characteristic
energy output of GRBs is $\sim 10^{52}~{\rm erg}$ and is
isotropic. The figure implies that at any time there should be of
order $\sim 15$ GRBs with redshifts $z\ga 5$ across the sky which are
brighter than $\sim 100$ nJy at an observed wavelength of $\sim
2\mu$m.  The infrared spectrum of these sources could be measured with
{\it NGST}\, as a follow-up on their early X-ray localization with
$\gamma$-ray or X-ray detectors. Prior to reionization, the spectrum
of GRB afterglows could reveal the long sought-after Gunn-Peterson
trough (Gunn \& Peterson 1965) due to absorption by the neutral
intergalactic medium.

\noindent
  
\noindent
  
The predicted GRB rate and flux are subject to uncertainties regarding
the beaming of the emission. The beaming angle may vary with observed
time due to the decline with time of the Lorentz factor $\gamma(t)$ of
the emitting material. As long as the Lorentz factor is significantly
larger than the inverse of the beaming angle (i.e., $\gamma \ga
\theta^{-1}$), the afterglow flux behaves as if it were emitted by a
spherically-symmetric fireball with the same explosion energy per unit
solid angle.  However, the lightcurve changes as soon as $\gamma$
declines below $\theta^{-1}$, due to the lateral expansion of the jet
(Rhoads 1997, 1999a,b; Panaitescu \& Meszaros 1999).  Finally, the
isotropization of the energy ends when the expansion becomes
sub-relativistic, at which point the remnant recovers the
spherically-symmetric Sedov-Taylor evolution for the total energy
output.  When $\gamma\sim 1$, the emission occurs from a roughly
spherical fireball with the effective explosion energy per solid angle
reduced by a factor of $(2\pi \theta^2/4\pi)$ relative to that at
early times, representing the fraction of sky around the GRB source
which is illuminated by the initial two (opposing) jets of angular
radius $\theta$ (see Ciardi \& Loeb 2000 for the impact of this
effect on the number counts). The calibration of the GRB event rate
per comoving volume, based on the number counts of GRBs (Wijers et
al.\ 1998), is inversely proportional to this factor.

The main difficulty in using GRBs as probes of the high-redshift
universe is that they are rare, and hence their detection requires
surveys which cover a wide area of the sky.  The simplest strategy for
identifying high-redshift afterglows is through all-sky surveys in the
$\gamma$-ray or X-ray regimes.  In particular, detection of
high-redshift sources will become feasible with the high trigger rate
provided by the forthcoming {\it Swift}\, satellite, to be launched in
2003 (see http://swift.gsfc.nasa.gov/, for more details).  {\it
Swift}\, is expected to localize $\sim$300 GRBs per year, and to
re-point within 20--70 seconds its on-board X-ray and UV-optical
instrumentation for continued afterglow studies. The high-resolution
GRB coordinates obtained by {\it Swift}\, will be transmitted to Earth
within $\sim$50 seconds. Deep follow-up observations will then be
feasible from the ground or using the highly-sensitive infrared
instruments on board {\it NGST}.\, {\it Swift}\, will be sufficiently
sensitive to trigger on the $\gamma$-ray emission from GRBs at
redshifts $z\ga 10$ (Lamb \& Reichart 2000).

%<>*************************************************************************

%<>
\section{OBSERVATIONAL PROBES OF THE EPOCH OF REIONIZATION}
\label{sec9}

%<>
\subsection{Inferring the Reionization Redshift Using \lya Photons}
\label{sec9.1.1}

The scattering cross-section of the \lya resonance line by neutral
hydrogen is given in Peebles (1993, \S 23). We consider a source at a
redshift $z_s$ beyond the redshift of reionization\footnote{We define
the reionization redshift to be the redshift at which the individual
\ionAR{H}{II} regions overlapped and most of the IGM volume was
ionized. In most realistic scenarios, this transition occurs rapidly
on a timescale much shorter than the age of the universe (see \S
\ref{sec6.3.1}). This is mainly due to the short distances between
neighboring sources.} $z_{i}$, and the corresponding scattering optical
depth of a uniform, neutral IGM between the source and the
reionization redshift. The optical depth is in general given by
integrating the absorption along the line of sight. At wavelengths
longer than \lya at the source, the optical depth obtains a small
value; these photons redshift away from the line center along its red
wing and never resonate with the line core on their way to the
observer.  Considering only this regime of the absorption wing leads
to an analytical result for the red damping wing of the Gunn-Peterson
trough (Miralda-Escud\'e 1998).

\begin{figure} % fig8.1
\psfig{file=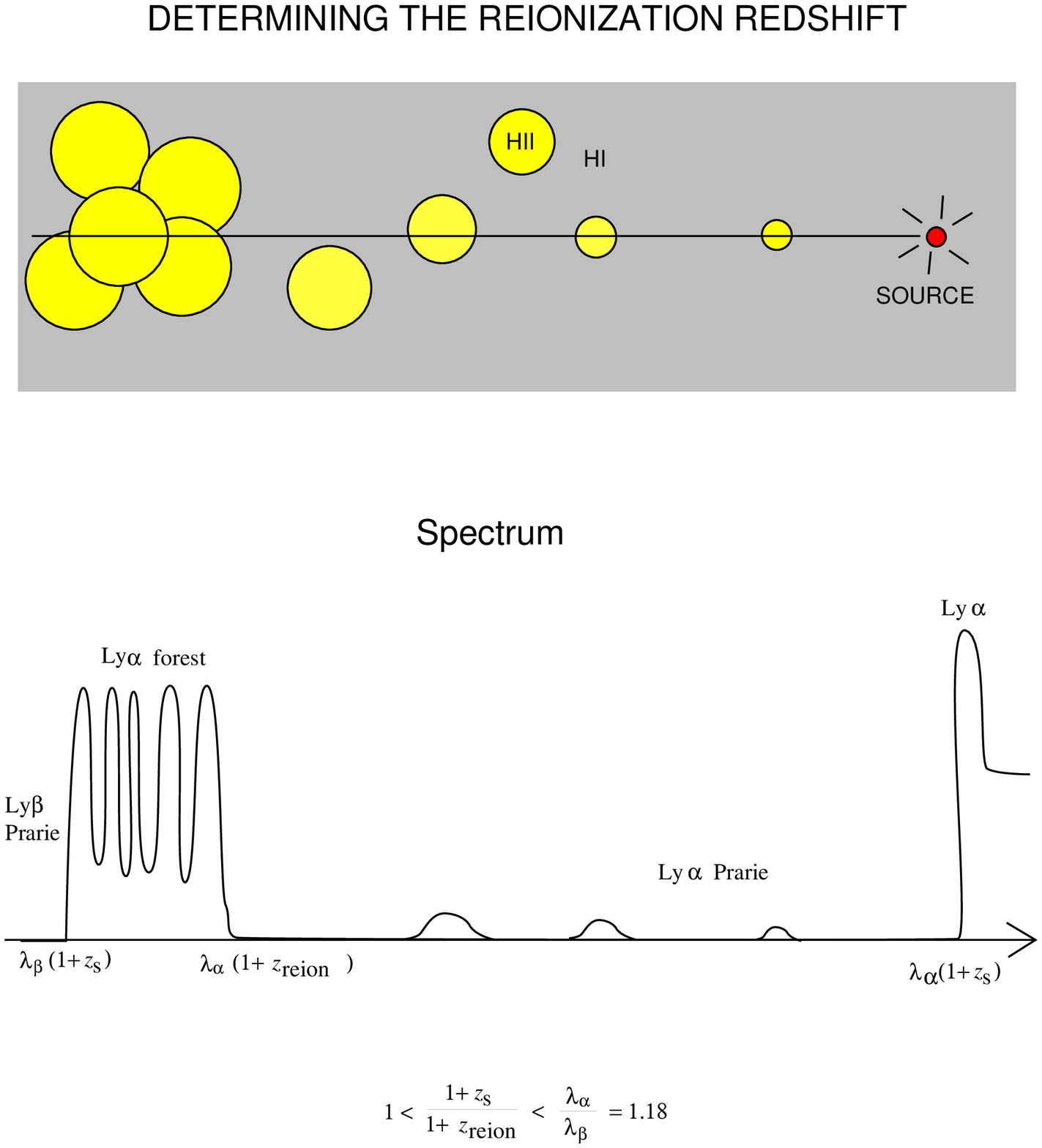,width=4.8in}
%\epsfbox{VIII_fig1.eps}
\caption{Sketch of the expected spectrum of a source at a redshift
$z_{s}$ slightly above the reionization redshift $z_{i}$. The
transmitted flux due to \ionAR{H}{II} bubbles in the pre-reionization era
and the Ly$\alpha$ forest in the post-reionization era is exaggerated
for illustration.}
\label{fig8a}
\end{figure}

At wavelengths corresponding to the \lya resonance between the source
redshift and the reionization redshift, $(1+z_{i})\lambda_\alpha\leq
\lambda_{\rm obs}\leq (1+z_s)\lambda_\alpha$, the optical depth is
given by the standard formula of Gunn \& Peterson (1965). Since this
optical depth is $\sim 10^5$, the flux from the source is entirely
suppressed in this regime. Similarly, the Ly$\beta$ resonance produces
another trough at wavelengths $(1+z_{i})\lambda_\beta\leq \lambda\leq
(1+z_s)\lambda_\beta$, where $\lambda_\beta=(27/32)\lambda_\alpha=
1026\,$\AA, and the same applies to the higher Lyman series lines. If
$(1+z_s)\geq 1.18(1+z_{i})$ then the \lya and the Ly$\beta$ resonances
overlap and no flux is transmitted in-between the two troughs (see
Figure~\ref{fig8a}). The same holds for the higher Lyman-series
resonances down to the Lyman limit wavelength of $\lambda_c=912$\AA.

At wavelengths shorter than $\lambda_c$, the photons are absorbed when
they photoionize atoms of hydrogen or helium. The bound-free
absorption cross-section from the ground state of a hydrogenic ion
(including neutral hydrogen singly-ionized helium) is given by
Osterbrock (1974). The cross-section for neutral helium is more
complicated; when averaged over its narrow resonances it can be fitted
to an accuracy of a few percent up to $h\nu=50$ keV by a simple
fitting function (Verner et al.\ 1986).

For rough estimates, the average photoionization cross-section for a
mixture of hydrogen and helium with cosmic abundances can be
approximated in the range of $54<h\nu \la 10^3$ eV as
$\sigma_{bf}\approx \sigma_0 (\nu/\nu_{\rm H,0})^{-3}$, where
$\sigma_0\approx 6\times 10^{-17}~{\rm cm^2}$ (Miralda-Escud\'e 2000).
For a source beyond the redshift of reionization, the resulting
bound-free optical depth only becomes of order unity in the extreme UV
to soft X-rays, around $h\nu \sim 0.1$ keV, a regime which is
unfortunately difficult to observe due to Galactic absorption
(Miralda-Escud\'e 2000).

A sketch of the overall spectrum of a source slightly above the
reionization redshift, i.e., with $1<[(1+ z_{\rm s})/(1+z_{i})]<1.18$,
is shown in Figure~\ref{fig8a}. The transmitted flux between the
Gunn-Peterson troughs due to Ly$\alpha$ and Ly$\beta$ absorption is
suppressed by the Ly$\alpha$ forest in the post-reionization
epoch. Transmission of flux due to \ionAR{H}{II} bubbles in the
pre-reionization epoch is expected to be negligible (Miralda-Escud\'e
1998). The redshift of reionization can be inferred in principle from
the spectral shape of the red damping wing (Miralda-Escud\'e \& Rees
1998; Miralda-Escud\'e 1998) or from the transmitted flux between the
Lyman series lines (Haiman \& Loeb 1999a).  However, these signatures
are complicated in reality by damped \lya systems along the
line-of-sight or by the inhomogeneity or peculiar velocity field of
the IGM in the vicinity of the source.  Moreover, bright sources, such
as quasars, tend to ionize their surrounding environment (Wood \& Loeb
2000) and the resulting \ionAR{H}{II} region in the IGM could shift the
\lya trough substantially (Cen \& Haiman 2000; Madau \& Rees 2000).

The inference of the \lya transmission properties of the IGM from the
observed spectrum of high-redshift sources suffers from uncertainties
about the precise emission spectrum of these sources, and in
particular the shape of their \lya emission line. The first galaxies
and quasars are expected to have pronounced recombination lines of
hydrogen and helium due to the lack of dust in their interstellar
medium (Oh 1999; Tumlinson \& Shull 2000; see also Baltz, Gnedin \&
Silk 1998). Lines such as $H_\alpha$ or the \ionAR{He}{II} 1640\,\AA\, 
line should reach the observer unaffected by the intervening IGM,
since their wavelength is longer than that of the Ly$\alpha$
transition which dominates the IGM opacity (Oh 1999).  However, as
described above, the situation is different for the \lya line photons.
As long as $z_s>z_i$, the intervening neutral IGM acts like a fog and
obscures the view of the \lya source [in contrast to the situation
with sources at $z_s<z_i$, where most of the intervening IGM is
ionized and only the blue wing of the \lya line is suppressed by the
Ly$\alpha$ forest (see Figure~\ref{fig1c})].  Photons which are
emitted at the \lya line center have an initial scattering optical
depth of $\sim 10^5$ in the surrounding medium.

\begin{figure}%fig8.3
\psfig{file=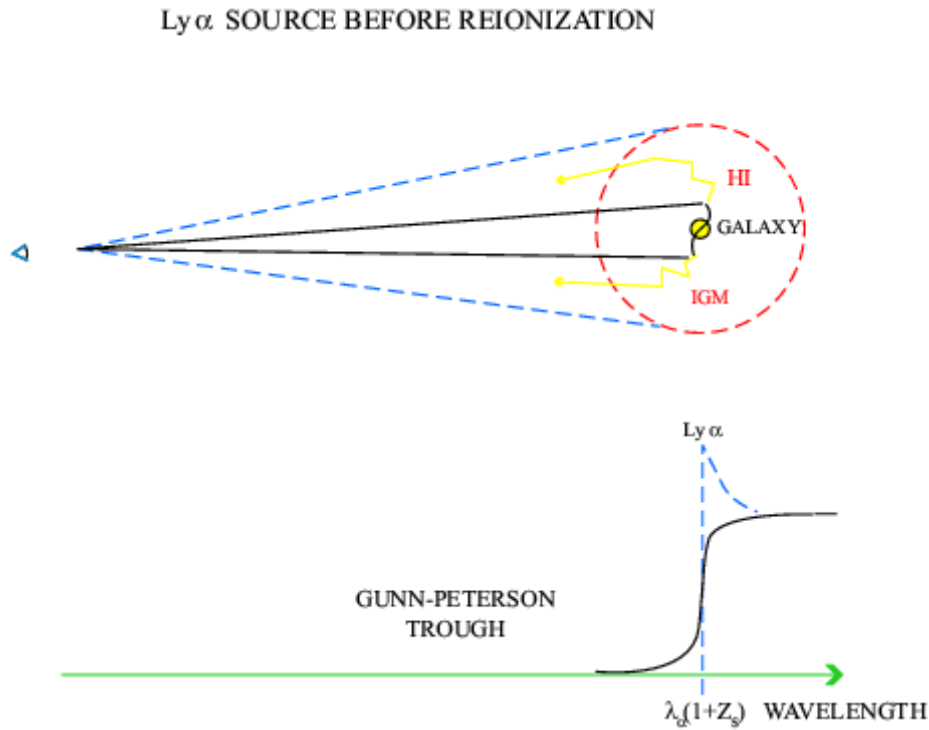,width=4.8in}
%\epsfbox{VIII_fig2.eps}
\caption{{\it Loeb-Rybicki halos:}\/ Scattering of \lya line photons
from a galaxy embedded in the neutral intergalactic medium prior to
reionization. The line photons diffuse in frequency due to the Hubble
expansion of the surrounding medium and eventually redshift out of
resonance and escape to infinity.  A distant observer sees a \lya halo
surrounding the source, along with a characteristically asymmetric
line profile. The observed line should be broadened and redshifted by
about one thousand ${\rm km~s^{-1}}$ relative to other lines (such as
H$_\alpha$) emitted by the galaxy.  }
\label{fig8c}
\end{figure}

The Ly$\alpha$ line photons are not destroyed but instead are absorbed
and re-emitted\footnote{At the redshifts of interest, $z_s\sim 10$,
the low densities and lack of chemical enrichment of the IGM make the
destruction of \lya photons by two-photon decay or dust absorption
unimportant.}.  Due to the Hubble expansion of the IGM around the
source, the frequency of the photons is slightly shifted by the
Doppler effect in each scattering event. As a result, the photons
diffuse in frequency to the red side of the \lya resonance.
Eventually, when their net frequency redshift is sufficiently large,
they escape and travel freely towards the observer (see
figure~\ref{fig8c}).  As a result, the source creates a faint
Ly$\alpha$ halo on the sky\footnote{The photons absorbed in the
Gunn-Peterson trough are also re-emitted by the IGM around the
source. However, since these photons originate on the blue side of the
\lya resonance, they travel a longer distance from the source,
compared to the \lya line photons, before they escape to the
observer. The Gunn-Peterson photons are therefore scattered from a
larger and hence dimmer halo around the source.  The Gunn-Peterson
halo is made even dimmer relative to the \lya line halo by the fact
that the luminosity of the source per unit frequency is often much
lower in the continuum than in the \lya line.}.  The well-defined
radiative transfer problem of a point source of \lya photons embedded
in a uniform neutral IGM was solved by Loeb \& Rybicki (1999). The
\lya halo can be simply characterized by the frequency redshift
relative to the line center, $(\nu-\nu_\alpha)$, which is required in
order to make the optical depth from the source equal to unity. At
$z_s \sim 10$, this frequency shift of $\nu_\star \sim 1\times
10^{13}$ Hz relative to the line center corresponds to a Doppler
velocity of $v\sim 10^3~{\rm km~s^{-1}}$. The halo size is then
defined by the corresponding proper distance from the source at which
the Hubble velocity provides a Doppler shift of this magnitude,
$r_\star \sim 1~{\rm Mpc}$.  Typically, the \lya halo of a source at
$z_{\rm s}\sim 10$ occupies an angular radius of $\sim
15^{\prime\prime}$ on the sky and yields an asymmetric line profile as
shown in Figure~\ref{fig8c}. The scattered photons are highly
polarized and so the shape of the halo would be different if viewed
through a polarization filter (Rybicki \& Loeb 1999).
%Figure 4 shows the intensity distribution of the
%halo in an unpolarized (left) and polarized filters (right).

\noindent
  
Detection of the diffuse \lya halos around bright high redshift
sources (which are sufficiently rare so that their halos do not
overlap) would provide a unique tool for probing the distribution and
the velocity field of the neutral intergalactic medium before the
epoch of reionization.  The \lya sources serve as lamp posts which
illuminate the surrounding \ionAR{H}{I} fog. On sufficiently large
scales where the Hubble flow is smooth and the gas is neutral, the
Ly$\alpha$ brightness distribution can be used to determine the
cosmological mass densities of baryons and matter. Due to their low
surface brightness, the detection of \lya halos through a narrow-band
filter is much more challenging than direct observation of their
sources at somewhat longer wavelengths. However, \NGST\, might be able
to detect the \lya halos around sources as bright as the quasar
discovered by Fan et al.\ (2000) at $z=5.8$ or the galaxy discovered by
Hu et al.\ (1999) at $z=5.74$, even if these sources were moved out to
$z\sim 10$ (see \S 4 in Loeb \& Rybicki 1999). The disappearance of
\lya halos below a certain redshift can be used to determine $z_{i}$.

\subsection{21 cm Tomography Of The Reionization Epoch}
\label{sec9.1.2}

The ground state of hydrogen exhibits hyperfine spitting involving the
spins of the proton and the electron. The state with parallel spins (the
triplet state) has a slightly higher energy than the state with
anti-parallel spins (the singlet state). The 21 cm line associated with the
spin-flip transition from the triplet to the singlet state is often used to
detect neutral hydrogen in the local universe. At high redshift, the
occurrence of a neutral pre-reionization IGM offers the prospect of
detecting the first sources of radiation and probing the reionization era
by mapping the 21 cm emission from neutral regions. While its energy
density is estimated to be only a $1\%$ correction to that of the CMB, the
redshifted 21 cm emission should display angular structure as well as
frequency structure due to inhomogeneities in the gas density field (Hogan
\& Rees 1979; Scott \& Rees 1990), hydrogen ionized fraction, and spin
temperature (Madau, Meiksin, \& Rees 1997). Some of the resulting
signatures during the pre-overlap phase of reionization (\S \ref{sec6.3.1})
and during the overlap phase are discussed by Tozzi et al.\ (2000) and
Shaver et al.\ (1999), respectively. Also, the 21 cm signatures have been
explored in a numerical simulation by Gnedin \& Ostriker (1997). Indeed, a
detailed mapping of \ionAR{H}{II} regions as a function of redshift is
possible in principle, and should be within the reach of proposed
instruments such as the Square Kilometer Array (hereafter SKA; see Taylor
\& Braun 1999). Serious technical challenges and foreground contamination
must, however, be overcome even for an initial detection of the
reionization signal.

The basic physics of the hydrogen spin transition is determined as
follows (for a more detailed treatment, see Madau et al.\ 1997). The
ground-state hyperfine levels of hydrogen tend to thermalize with the
CMB background, making the IGM unobservable. If other processes shift
the hyperfine level populations away from thermal equilibrium, then
the gas becomes observable against the CMB in emission or in
absorption. The energy of the 21 cm transition is $E_{21}=5.9 \times
10^{-6}$ eV, corresponding to a frequency of 1420 MHz. In the presence
of the CMB alone, the spin states reach thermal equilibrium on a
timescale which is much shorter than the age of the universe at all
redshifts after cosmological recombination.

The IGM is observable when the kinetic temperature of the gas differs
from $T_{\rm CMB}$ and an effective mechanism couples the hyperfine
levels to the gas temperature. Although collisional de-excitation of
the triplet level (Purcell \& Field 1956) is a possible mechanism, in
the low-density IGM the dominant mechanism is scattering by Ly$\alpha$
photons (Wouthuysen 1952; Field 1958). Continuum UV photons produced
by early radiation sources redshift by the Hubble expansion into the
local Ly$\alpha$ line at a lower redshift. These photons mix the spin
states via the Wouthuysen-Field process whereby an atom initially in
the $n=1$ state absorbs a Ly$\alpha$ photon, and the spontaneous decay
which returns it from $n=2$ to $n=1$ can result in a final spin state
which is different from the initial one. Efficient thermalization of
the hyperfine levels occurs above a particular critical scattering
rate (Madau et al.\ 1997).

Assuming such efficient thermalization, a heated patch of neutral
hydrogen at the mean density is observed at a differential antenna
temperature $\delta T_b$ relative to the CMB given by Madau et al.\
(1997). In overdense regions, the observed $\delta T_b$ is
proportional to the overdensity, and in partially ionized regions
$\delta T_b$ is proportional to the neutral fraction. At high
temperatures, the IGM is observed in emission at a level ($\delta T_b
\sim 28\, {\rm mK}$) that is independent of temperature. On the other
hand, a cool IGM is observed in absorption at a level that increases
with temperature and becomes larger than the maximum $\delta T_b$
observable in emission. As a result, a number of cosmic events are
expected to leave observable signatures in the redshifted 21 cm line.

Since the CMB temperature is only $2.73(1+z)$ K, even relatively
inefficient heating mechanisms are expected to heat the IGM above
$T_{\rm CMB}$ well before reionization. Possible preheating sources
include soft X-rays from early quasars or star-forming regions, as
well as thermal bremsstrahlung from ionized gas in collapsing
halos. However, even the radiation from the first stars may suffice
for an early preheating. Only $\sim 10\%$ of the present-day global
star formation rate is required (Madau et al.\ 1997) for a
sufficiently strong Ly$\alpha$ background which produces a scattering
rate above the thermalization rate. Such a background produces
absorption, since the kinetic gas temperature is initially lower than
$T_{\rm CMB}$ because of adiabatic expansion. Thus, the entire IGM can
be seen in absorption, but the IGM is then heated above $T_{\rm CMB}$
in $\sim 10^8$ yr (Madau et al.\ 1997) by the atomic recoil in the
repeated resonant Ly$\alpha$ scattering. According to \S \ref{sec8.1}
(also compare Gnedin 2000a), the required level of star formation is
expected to be reached already at $z \sim 20$, with the entire IGM
heated well above the CMB by the time overlap begins at $z \sim
10$. Thus, although the initial absorption signal is in principle
detectable with the SKA (Tozzi et al.\ 2000), it likely occurs at $\la
100$ MHz where Earth-based radio interference is highly problematic.

As individual ionizing sources turn on in the pre-overlap stage of
reionization, the resulting \ionAR{H}{II} bubbles may be individually
detectable if they are produced by rare and luminous sources such as
quasars. If the \ionAR{H}{II} region expands into an otherwise
unperturbed IGM, then the expanding shell can be mapped as follows
(Tozzi et al.\ 2000). The \ionAR{H}{II} region itself, of course,
shows neither emission nor absorption. Outside the ionized bubble, a
thin shell of neutral gas is heated above the CMB temperature and
shows up in emission. A much thicker outer shell is cooler than the
CMB due to adiabatic expansion, and it produces absorption. Finally,
at large distances from the quasar, the gas approaches $T_{\rm CMB}$ as
the quasar radiation weakens. For a quasar with an ionizing intensity
of $10^{57}$ photons s$^{-1}$ observed after $\sim 10^7$ yr with
$2\arcmin$ resolution and 1 MHz bandwidth, the signal ranges from -3
to 3 $\mu$Jy per beam (Tozzi et al.\ 2000). Mapping such regions would
convey information on the quasar number density, ionizing intensity,
opening angle, and on the density distribution in the surrounding
IGM. Note, however, that an \ionAR{H}{II} region which forms at a
redshift approaching overlap expands into a preheated IGM. In this
case, the \ionAR{H}{II} region itself still appears as a hole in an
otherwise emitting medium, but the quasar-induced heating is not
probed, and there is no surrounding region of absorption to supply an
enhanced contrast.

At redshifts approaching overlap, the IGM should be almost entirely
neutral but also much hotter than the CMB.. In this redshift range
there should still be an interesting signal due to density
fluctuations. The same cosmic network of sheets and filaments that
gives rise to the Ly$\alpha$ forest observed at $z \la 5$ should lead
to fluctuations in the 21 cm brightness temperature at higher
redshifts. At 150 MHz ($z=8.5$), for observations with a bandwidth of
1 MHz, the root mean square fluctuation should be $\sim 10$ mK at $1
\arcmin$, decreasing with scale (Tozzi et al.\ 2000).

A further signature, observable over the entire sky, should mark the
overlap stage of reionization. During overlap, the IGM is transformed
from being a neutral, preheated and thus emitting gas, to being almost
completely ionized. This disappearance of the emission over a
relatively narrow redshift range can be observed as a drop in the
brightness temperature at the frequencies corresponding to the latter
stages of overlap (Shaver et al.\ 1999). This exciting possibility,
along with those mentioned above, face serious challenges in terms of
signal contamination and calibration. The noise sources include
galactic and extragalactic emission sources, as well as terrestrial
interference, and all of these foregrounds must be modeled and
accurately removed in order to observe the fainter cosmological signal
(see Shaver et al.\ 1999 for a detailed discussion). For the overlap
stage in particular, the sharpness of the spectral feature is the key
to its detectability, but it may be significantly smoothed by
inhomogeneities in the IGM.

%<>
\subsection{Effect of Reionization on CMB Anisotropies}
\label{sec9.2}

In standard cosmological models, the early universe was hot and
permeated by a nearly uniform radiation bath. At $z \sim 1200$ the
free protons and electrons recombined to form hydrogen atoms, and most
of the photons last scattered as the scattering cross-section dropped
precipitously. These photons, observed today as the Cosmic Microwave
Background (CMB), thus yield a snapshot of the state of the universe
at that early time. Small fluctuations in the density, velocity, and
gravitational potential lead to observed anisotropies (e.g., Bennett
et al.\ 1996) that can be analyzed to yield a great wealth of
information on the matter content of the universe and on the values of
the cosmological parameters (e.g., Hu 1995).

Reionization can alter the anisotropy spectrum, by erasing some of the
primary anisotropy imprinted at recombination, and by generating
additional secondary fluctuations that could be used to probe the era
of reionization itself (see Haiman \& Knox 1999 for a review). The
primary anisotropy is damped since the rescattering leads to a
blending of photons from initially different lines of sight.
Furthermore, not all the photons scatter at the same time, rather the
last scattering surface has a finite thickness. Perturbations on
scales smaller than this thickness are damped since photons scattering
across many wavelengths give canceling redshifts and blueshifts. If
reionization occurs very early, the high electron density produces
efficient scattering, and perturbations are damped on all angular
scales except for the very largest.

The optical depth to scattering out to some redshift $z$ depends
simply on the cosmological parameters (e.g., Hu 1995). With our
standard parameters ($\Omm=0.3$, $\Oml=0.7$, $\Omega_b=0.045$, and
$h=0.7$) this implies $\tau=0.037$ at the current lower limit on
reionization of $z=5.8$ (Fan et al.\ 2000), with $\tau=0.10$ if
$z=11.6$ and $\tau=0.15$ if $z=15.3$. Recent observations of
small-scale anisotropies (Lange et al.\ 2000; Balbi et al.\ 2000)
revealed a peak in the power spectrum on a $\sim 1^\circ$ scale, as
expected from the primary anisotropies in standard cosmological
models. This indicates that the reionization damping, if present, is
not very large, and the observations set a limit of $\tau < 0.33$ at
$95\%$ confidence (Tegmark \& Zaldarriaga 2000) and, therefore, imply
that reionization must have occurred at $z \la 30$.

However, measuring a small $\tau$ from the temperature anisotropies
alone is expected to be very difficult since the anisotropy spectrum
depends on a large number of other parameters, creating a
near-degeneracy which limits our ability to measure each parameter
separately. However, Thomson scattering also creates net polarization
for incident radiation which has a quadrupole anisotropy. This
anisotropy was significant at reionization due to large-scale
structure which had already affected the gas distribution. The result
is a peak in the polarization power spectrum on large angular scales
(tens of degrees). Although experiments must overcome systematic
errors from the detector itself and from polarized foregrounds (such
as galactic dust emission and synchrotron radiation), parameter
estimation models (Eisenstein, Hu, \& Tegmark 1999; Zaldarriaga,
Spergel, \& Seljak 1997) suggest that the peak can be used to measure
even very small values of $\tau$: $2\%$ for the upcoming MAP
satellite, and $0.5\%$ for the Planck satellite which will reach
smaller angular scales with higher accuracy.

Reionization should also produce additional temperature anisotropies
on small scales. These result from the Doppler effect. By the time of
reionization, the baryons have begun to follow dark matter potentials
and have acquired a bulk velocity. Since the electrons move with
respect to the radiation background, photons are given a Doppler kick
when they scatter off the electrons. Sunyaev (1978) and Kaiser (1984)
showed, however, that a severe cancellation occurs if the electron
density is homogeneous. Opposite Doppler shifts on crests and troughs
of a velocity perturbation combine to suppress the anisotropy induced
by small-scale velocity perturbations. The cancellation is made more
severe by the irrotational nature of gravitationally-induced flows.
However, if the electron density varies spatially, then the scattering
probability is not equal on the crest and on the trough, and the two
do not completely cancel. Since a positive effect requires variation
in both electron density and velocity, it is referred to as a
second-order anisotropy.

The electron density can vary due to a spatial variation in either the
baryon density or the ionized fraction. The former is referred to as
the Ostriker-Vishniac effect (Ostriker \& Vishniac 1986; Vishniac
1987). The latter depends on the inhomogeneous topology of
reionization, in particular on the size of \ionAR{H}{II} regions due to
individual sources (\S \ref{sec6.2}) and on spatial correlations among
different regions. Simple models have been used to investigate the
character of anisotropies generated during reionization (Gruzinov \&
Hu 1998; Knox et al.\ 1998; Aghanim et al.\ 1996). The
Ostriker-Vishniac effect is expected to dominate all anisotropies at
small angular scales (e.g., Jaffe \& Kamionkowsky 1998), below a tenth
of a degree, because the primary anisotropies are damped on such small
scales by diffusion (Silk damping) and by the finite thickness of the
last scattering surface. Anisotropies generated by inhomogeneous
reionization may be comparable to the Ostriker-Vishniac effect, and
could be detected by MAP and Planck, if reionization is caused by
bright quasars with 10 Mpc-size ionized bubbles. However, the smaller
bubbles expected for mini-quasars or for star-forming dwarf galaxies
would produce a weaker anisotropy signal at smaller angular scales,
likely outside the range of the upcoming satellites (see, e.g., Haiman
\& Knox 1999 for discussion). Gnedin \& Jaffe (2000) used a numerical
simulation to show that, in the case of stellar reionization, the
effect on the CMB of patchy reionization is indeed sub-dominant
compared to the contribution of non-linear density and velocity
fluctuations. Nevertheless, a signature of reionization could still be
detected in future measurements of CMB angular fluctuations on the
scale of a few arcseconds.

%<>***************************************************************************

%<>
\section{CHALLENGES FOR THE FUTURE}
\label{sec10}

{\it When and how did the first stars and black holes form and when
and how did they ionize most of the gas in the universe?} In this
review we have sketched the first attempts to answer these questions
and the basic physical principles that underlie these attempts. The
coming decade will likely be marked by major advances in our ability
to make theoretical predictions in an attempt to answer these
questions, and will culminate with the launch of {\it NGST},\, a
telescope which is ideally suited for testing these predictions. At
about the same time, the Planck satellite (and perhaps MAP before it)
is expected to directly infer the reionization redshift from
measurements of the CMB polarization power spectrum on large angular
scales. Also in about a decade, next-generation arrays of radio
telescopes may detect the 21 cm emission from the pre-reionization,
neutral warm IGM. The difficult questions just mentioned will receive
their ultimate answers from observations, but it surely is fun to try
to find the answers theoretically in advance, before we can deduce
them by looking through our most technologically-advanced telescopes.

%<>***************************************************************************
%%\acknowledgements

\vspace{.2in}

AL thanks the Institute for Advanced Study at Princeton for its kind
hospitality when the writing of this review began. RB acknowledges
support from Institute Funds; support by the Smithsonian Institution
Visitor Program during a visit to the Harvard-Smithsonian CfA; and the
hospitality of the Weizmann Institute, Israel, where part of this
review was written. This work was supported in part by NASA grants NAG
5-7039, 5-7768, and NSF grants AST-9900877, AST-0071019 for AL.

%\vfill\eject
\vspace{.5in}
%<>***************************************************************************
%***
%***references
%***

%***************************************************************************

\end{document}